\documentclass[useAMS,usenatbib]{mn2e}
\usepackage{times,graphicx,rotating}
\usepackage{longtable,supertabular,multirow}
\usepackage{amsmath,amssymb}

\newcommand{\mnras}{MNRAS}
\newcommand{\apj}{ApJ}
\newcommand{\apjs}{ApJS}
\newcommand{\apjl}{ApJ}
\newcommand{\aj}{AJ}
\newcommand{\araa}{ARA\&A}
\newcommand{\aap}{A\&A}
\newcommand{\aaps}{A\&AS}

\newcommand{\pasa}{PASA}
\newcommand{\jqsrt}{JQS\&RT}

\newcommand{\cmmthree}{\mbox{cm$^{-3}$}}

\newcommand{\kms}{\mbox{km\,s$^{-1}$}}
\newcommand{\degree}{\mbox{$^{\circ}$}}

\newcommand{\chthreecn}{\mbox{CH$_3$CN}}
\newcommand{\chthreeoh}{\mbox{CH$_{3}$OH}}
\newcommand{\hcop}{\mbox{HCO$^+$}}
\newcommand{\hthirteencop}{\mbox{H$^{13}$CO$^+$}}

\newcommand{\rarr}{\rightarrow}

\title[A \chthreecn~and \hcop~survey towards southern methanol masers]{A \chthreecn~and \hcop~survey
  towards southern methanol masers associated with star formation}
\author[C. R. Purcell {\it et al.}]{C. R. Purcell$^{1}$,
  R. Balasubramanyam$^{1,2}$, 
  M. G. Burton$^{1}$, 
  A. J. Walsh$^{1}$,
  V. Minier$^{1,3,4}$, \newauthor 
  M. R. Hunt-Cunningham$^{1}$,
  L. L. Kedziora-Chudczer$^{1,6}$, 
  S. N. Longmore$^{1,5}$,
  T. Hill$^{1}$, \newauthor 
  I. Bains$^{1}$, 
  P. J. Barnes$^{1,6}$,
  A. L. Busfield$^{7}$, 
  P. Calisse$^{1,8}$,
  N. H. M. Crighton$^{1}$, \newauthor
  S. J. Curran$^{1}$,
  T. M. Davis$^{1,9}$,
  J. T. Dempsey$^{1}$, 
  G. Derragopian$^{1}$, 
  B. Fulton$^{1,10}$, \newauthor
  M. G. Hidas$^{1}$, 
  M. G. Hoare$^{7}$,
  J.-K. Lee$^{1,11}$, 
  E. F. Ladd$^{12,1,5}$, 
  S. L. Lumsden$^{7}$, \newauthor
  T. J. T. Moore$^{13}$, 
  M. T. Murphy$^{1,14}$, 
  R. D. Oudmaijer$^{6}$, 
  M. B. Pracy$^{1}$, \newauthor 
  J. Rathborne$^{1,15}$,
  S. Robertson$^{5}$, 
  A. S. B. Schultz$^{1}$, 
  J. Shobbrook$^{1}$,
  P. A. Sparks$^{1}$, \newauthor 
  J. Storey$^{1}$, 
  T. Travouillion$^{1,16}$\\
$^{1}$~~School of Physics, University of New South Wales, Sydney, NSW 2052,
  Australia\\
$^{2}$~~Raman Research Institute, Sadashivanagar, Bangalore 560 080,
  India\\
$^{3}$~~Service d'Astrophysique, DAPNIA/DSM/CEA Saclay, 91191
  Gif-sur-Yvette, France\\
$^{4}$~~AIM, Unit\'e Mixte de Recherche, CEA$-$CNRS$-$Universit\'e Paris VII, UMR 7158, CEA/Saclay, 91191 Gif sur Yvette, France\\
$^{5}$~~CSIRO Australia Telescope National Facillity, PO Box 76, Epping,
NSW 1710, Australia\\
$^{6}$~~School of Physics, University of Sydney, NSW 2006, Australia\\
$^{7}$~~School of Physics and Astronomy, University of Leeds, Leeds
  LS2 9JT, UK\\
$^{8}$~~School of Physics and Astronomy, University of Wales, Cardiff, 5, 
The Parade, Cardiff CF24 3YB, Wales, UK\\
$^{9}$~~Research School of Astronomy and Astrophysics, The Australian
National University,\\ ~~~Mount Stromlo Observatory, Cotter Road, Weston
Creek, ACT 2611, Australia\\
$^{10}$~~Centre for Astronomy, James Cook University, Townsville, QLD
  4811, Australia\\
$^{11}$~Department of Pure and Applied Physics, Queen's University, Belfast
BT7 1NN, U.K.\\
$^{12}$~Department of Physics, Bucknell University, Lewisburg, PA 17837,
USA\\
$^{13}$~Astrophysics Research Institute, Liverpool John Moores
University, 12 Quays House,\\ ~~~ Egerton Wharf, Birkenhead CH41 1LD, UK\\
$^{14}$~Institute of Astronomy, University of Cambridge, Madingley Road,
Cambridge CB3 OHA, UK\\
$^{15}$~Institute for Astrophysical Research, Boston University, Boston,
MA 02215, USA\\
$^{16}$~California Institute of Technology, 1200 E. California Blvd, Pasadena
91125, CA, USA}
\begin{document}

\pagerange{\pageref{firstpage}--\pageref{lastpage}} \pubyear{2005}

\maketitle

\label{firstpage}

\begin{abstract}
We present the initial results of a 3-mm spectral line survey towards
83 methanol maser selected massive star-forming regions. Here we
report observations of the J\,=\,5\,--\,4 and 6\,--\,5 rotational transitions of
methyl cyanide (\chthreecn) and the J\,=\,1\,--\,0 transition of \hcop~and
\hthirteencop.

\chthreecn~emission is detected in 58 sources (70\,\% of our
sample). We estimate the temperature and column density for 37 of
these using the rotational diagram method. The
temperatures we derive range from 28\,--\,166\,K, and are lower
than previously reported temperatures, derived from higher J
transitions. We find that \chthreecn~is brighter and more commonly
detected towards ultra-compact H{\scriptsize II} (UCH{\scriptsize II})
regions than towards isolated maser sources. Detection of
\chthreecn~towards  isolated maser sources strongly suggests that
these objects are internally heated and that \chthreecn~is excited
prior to the UCH{\scriptsize II} phase of massive star-formation.

\hcop~is detected towards 82 sources (99\,\% of our sample), many of
which exhibit 
asymmetric line profiles compared to \hthirteencop. Skewed profiles
are indicative of inward or outward motions, however, we find
approximately equal numbers of red and blue-skewed profiles among all
classes. Column densities are derived from an analysis of the
\hcop~and \hthirteencop~line profiles.

80 sources have mid-infrared counterparts: 68 seen in emission and
12 seen in absorption as `dark clouds'. Seven of the twelve dark
clouds exhibit  asymmetric \hcop~profiles, six of which are skewed to
the blue, indicating infalling motions. \chthreecn~is also common in
dark clouds, where it has a 90\,\% detection rate.
\end{abstract}



\section{Introduction}
Identifying young massive stars ($>$\,8\,M$_{\odot}$) in their early
evolutionary phases is an important step in attempting to understand
massive star formation and its effect on the Galactic ecology. 

Massive stars begin their lives at the centre of dense contracting
cores, embedded in giant molecular clouds (GMCs). At the earliest
stages of their evolution they heat the surrounding dust, and are
visible as luminous sub-millimetre (mm) and far-infrared sources
\citep{Kurtz2000}. High densities existing prior to the final collapse
cause simple organic molecules present in the interstellar medium to
freeze onto the dust  mantles. Grain surface chemistry leads to the
production of heavy ices such as methanol (\chthreeoh). As the
protostar evolves, heat and radiation evaporate these species,
expelling them into gas phase where they fuel a rich `hot core'
chemistry \citep{VanDishoeck1998}. Eventually, the embedded protostar
ionises the surrounding environment, forming a ultra-compact
H{\scriptsize II} (UCH{\scriptsize II}) region and complex molecules
are destroyed owing to continued exposure to heat and radiation.

Targeted surveys towards known star forming regions have suggested a
close relationship between class II methanol masers, hot cores and
UCH{\scriptsize II} regions \citep{Menten1991,Minier2001}. Parkes 64-m
and Australia Telescope Compact Array (ATCA) observations by
\citet{Walsh1998,Walsh1999} have shown that only 25\,\% of methanol
masers are directly associated with radio continuum emission, tracing
UCH{\scriptsize II} regions. Subsequent follow-up observations towards
the {\it isolated} masers confirm that all maser sites are
associated with luminous sub-mm continuum emission \citep{Walsh2003}.

In the present paper we investigate the link between methanol masers,
UCH{\scriptsize II} regions and the chemically active `hot molecular
core' (HMC) phase of massive star formation. We present the initial
results of a single-dish molecular line survey of 83 methanol maser
selected massive star forming regions.


\subsection{\chthreecn~Spectroscopy}
The primary chemical tracer in this survey is methyl cyanide
(\chthreecn). \chthreecn~is a good tracer of the conditions found in
HMCs owing to its favourable abundance and excitation in warm
($>$\,100\,K) and dense ($>$\,10$^5$\,\cmmthree) regions. It is
thought to form through reactions between species evaporated from dust
grain mantles (e.g. \citealt*{Millar1997b}) or through gas phase
chemistry in the envelope around massive young stars (see
\citealt*{Mackay1999}). Grain-surface chemistry has also been invoked
to explain the enhanced abundances observed in HMCs, however, in all
formation scenarios \chthreecn~is evident in observations after the
core temperature rises past $\sim$\,90\,K. 

\chthreecn~is a member of the ${\rm c_{3v}}$ group of symmetric tops,
whose rotational energy levels may be described by two quantum
numbers: J, the total angular momentum and K, the projection of J
along the axis of symmetry. Individual J\,$\rarr$\,(J\,--\,1)
transitions are grouped into `rotational ladders' labelled by their K
values (see \citet*{LorenMundy1984} for a detailed description). For
each J\,$\rarr$\,(J\,--\,1) transition, selection rules prohibit
radiative transitions between the K ladders and their relative
populations are determined exclusively by collisional
excitation. Assuming local thermal equilibrium (LTE) and optically
thin lines, the relative intensities of the K components yield a
direct measure of the kinetic temperature and column density. 

The energy spacings between individual J levels are almost independent
of K ladder, however, increasing centrifugal distortion causes
successive K components to shift to progressively lower
frequencies. The offset in frequency is slight, and the K components
of a particular J\,$\rarr$\,(J\,--\,1) transition may be observed
simultaneously in a single bandpass, minimising errors in their
relative calibration.  

Spin statistics of the hydrogen nuclei divide \chthreecn~into two spin
states, dubbed A and E. Energy levels with ${\rm
  K=3n,~n=0,1,2\,\ldots}$ belong to the A state, while those with
${\rm K\neq 3n,~n=0,1,2\,\ldots}$ belong to the E state. The A states
have twice the statistical weight of the E states. Neither radiation
nor collisions convert between states; if formed in equilibrium
conditions the A\,/\,E abundance ratio is expected to be $\sim$\,1
(\citealt{Minh1993}). 


\subsection{\hcop~Spectroscopy}
\hcop~and its isotopomer, \hthirteencop, were also observed in order
to probe the kinematics of the extended envelopes and as a diagnostic
of the optical depth towards the cores. A comparison between the
generally optically thick \hcop~line and the optically thin
\hthirteencop~line yields information on the bulk motions of gas in
the regions. 

\hcop~has a large dipole moment and is a highly abundant molecule,
with abundance especially enhanced around regions of higher fractional
ionisation. \hcop~is also enhanced by the presence of outflows where
shock generated radiation fields are present (\citet*{Rawlings2000};
\citet{Rawlings2004}). It shows saturated and self absorbed line
profiles around massive star-forming regions and is a good tracer of
the dynamics in the vicinity of young protostellar objects
(e.g. \citealt*{DeVries2005}). The emission from its isotopomer,
\hthirteencop, is generally optically thin and traces similar gas
densities to \hcop~at $n\sim 10^5$\,cm$^{-3}$.

 Table~\ref{tab:transitions} presents the details of the observed
 transitions. All molecular constants are in SI units and come from
 the catalog of \citet{Pickett1998}.


We justify the source selection criteria and derive the distances and
luminosities in the following section. In \S\,3 we describe the
observations and data reduction methodology. In \S\,4, we present the
initial results of the line survey and describe the data. We outline
the analysis techniques used and present derived physical parameters
in \S\,5, and in \S\,6 we analyse and discuss general trends in the
data and test the validity of our results. Finally, in \S\,7 we
conclude with a summary of our investigations and suggest future
avenues of investigation. The full set of spectra, MSX images and
related plots are presented as an online addition to this paper.

\section{Source Selection}
77 sources were drawn from the methanol maser, radio and
sub-millimetre surveys conducted by
\citet{Walsh1997,Walsh1999,Walsh2003}. The original sources were
chosen for their red IRAS colours using the colour selection criteria
of \citet{WoodChurchwell1989b}, to be indicative of UCH{\scriptsize
  II} regions. Our sub-sample of 6.71\,GHz methanol maser sites are
all associated with warm, dusty clumps, traced by sub-millimetre
continuum emission \citep{Walsh2003}. The final selection was also
constrained by the latitude of the Mopra antenna at -31\degree. We
required that the sources remain above an elevation of 30\degree~for
at least three hours between the months June to September so allowed
declinations were restricted between 0 and $-$60 degrees. The sample
are split into two main categories: eighteen maser sites are within a
Mopra beam (33\,\arcsec) of UCH{\scriptsize II} regions. The remaining
59 maser-sites have no detectable radio emission and are associated
only with thermal emission in the sub-mm and infrared. An additional
six `maserless cores', only detected in thermal emission, were added
to the list, bringing the total sample to 83. These cool dusty clumps
were selected from recent 1.2-mm continuum observations by
\citet{Hill2005} and are found in the same fields but offset from the
masers and radio emission. One such core was subsequently found to be
coincident with a UCH{\scriptsize II} region but the others are not
directly associated with other tracers of massive star formation. We
investigate the remaining five as potential precursors to the hot core
phase.

Table~\ref{tab:sources} presents the source details and their
associations. The first column is the abbreviated Galactic name,
columns two and three are the J2000 equatorial coordinates of the
pointing centre (usually coincident with the maser site). Column five
presents the adopted kinematic distances in kpc. The approximate
bolometric luminosity and single-star spectral type are noted in
columns six and seven and any associated methanol maser,
radio-continuum or mid-infrared thermal emission is noted in column
eight. Alternative names used in the literature are presented in the
final column.


\subsection{Kinematic Distances}
The majority of sources are in the first quadrant of the Galactic
plane, between longitude ($l$) 0\degree~and 35\degree.  Five sources
are located between $l$ = 316\degree~and 335\degree. We used the
rotation curve of \citet*{BrandBlitz1993} to calculate the kinematic
distance, assuming a distance to the Galactic centre of 8.5\,kpc and a
solar angular velocity of 220\,\kms. In most cases the V$_{\rm LSR}$
was determined from Gaussian fits to the \hthirteencop~line. 

All sources lie inside the solar circle and thus have a near\,/\,far
distance ambiguity. We were able to resolve the ambiguity for 58
sources, mainly using the work of \citet{Downes1980},
\citet{Solomon1987} and \citet*{Wink1982}. Also, where the near and
far kinematic distances are within $\sim$ 1 kpc we have assigned the
distance of the tangent point to the source. We favoured the near
distance when a source was seen in absorption in
8\,\micron~images. Several sources have been identified in the
literature as being part of the same complex and we have placed them
at a common distance. The remaining 25 sources are assumed to lie at
the near kinematic distance, except for G10.10+0.72, which was not
convincingly detected in any molecular tracer and was placed at the
far distance.

Assuming peculiar velocities $\pm$\,10\,\kms~from the model Galactic
rotation curve, we estimate the error in the distance is
0.65\,$\pm$\,0.29\,kpc, or 20\,$\pm$\,9\,\% of the near kinematic
distance. Eighteen sources in our sample lie between
$l$\,=\,0$^{\circ}$ and $l$\,=\,10$^{\circ}$. Over these longitudes
the Galactic rotation curve is not well determined and large peculiar
velocities will lead to correspondingly greater errors in the distance
determination.


\subsection{Bolometric luminosities}
If we assume the millimetre, sub-millimetre, far and near infrared
emission is associated with the same object, we can estimate a bolometric
luminosity by fitting a two-component greybody to the spectral energy
distribution (SED) (e.g. \citealt{Faundez2004,Minier2005} and
references therein). The model greybody SED is a summation of two
blackbodies, with separate temperatures and amplitudes, modified by a
common dust emissivity index, ${\rm \beta}$. Typical values for ${\rm
 \beta}$ range between one and two, with the Kramers-Kroing theorem
constraining the lower limit at one \citep{Emerson1988}. In the
literature a value of two is generally assumed
\citep{dunne2000,james2002,kramer2003} and recent results by Hill et
al. (in preparation) suggest an emmisivity of two for a sample of
massive star forming regions. Accordingly we limit ${\rm \beta}$
between values of one and two for our fits, while all other parameters
are free. No attempt was made to interpret the greybody fits, except
to measure the total intensity by integrating under the curve between
${\rm 10^6}$ and ${\rm 10^{16}}$\,Hz.
\begin{figure}
  \begin{center}
    \includegraphics[height=7.9cm, angle=-90, trim=0 0 0 0]{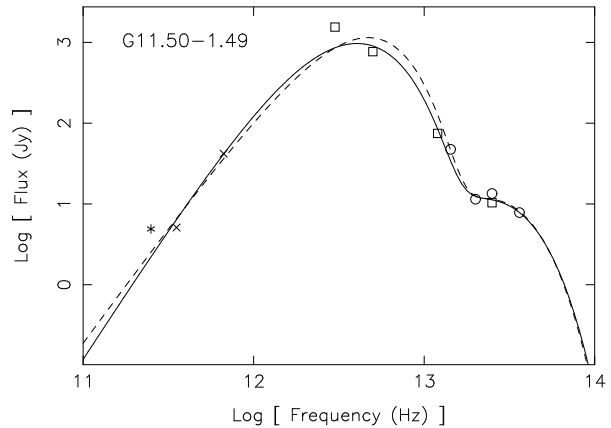}
    \caption{~Greybody fits to the SED of source
    G11.50$-$1.49. MSX data points are plotted as circles, IRAS data
    points are plotted as squares, SCUBA data points are plotted as
    crosses and SIMBA data points are plotted stars. The solid line
    represents a two-component fit to all data points, while the
    dashed line represents a fit to the SIMBA, SCUBA and MSX data
    only. Without the IRAS points the fit overestimates the integrated
    intensity by a factor of 1.5\,$\pm$\,0.8. The bolometric
    luminosity is estimated to be accurate to a within a factor of
    two.} 
    \label{fig:sed_example}
  \end{center}
\end{figure}

Sub-millimetre data for 78 maser sites comes from the observations of
\citet{Walsh2003} using the SCUBA bolometer on the James Clerk Maxwell
Telescope (JCMT). 1.2-mm fluxes were taken from the work of
\citet{Hill2005}. Far and mid-infrared fluxes are drawn from the
Infrared Astronomy Satellite (IRAS) point source catalog\footnote{IRAS
  images and photometry data are available online at
  http://www.ipac.gov.com} and measured directly from the Midcourse
Space Experiment (MSX) images. We have assumed 30\,\% errors on all
fluxes.

High source multiplicity in the Galactic plane means quoted IRAS
fluxes may contain contributions from several sources within the
1\,\arcmin~IRAS beam. We include IRAS data in the SED if the
source appears isolated in the higher resolution MSX image, or if the
MSX and IRAS fluxes match at 14\,\micron. In some cases the 60 and
100\,\micron~points are included as upper limits to the fit. Without
data points at $\sim$\,100\,\micron, the peak of the SED is not well
constrained. In an attempt to measure the effect of missing data, we
fit fifteen SEDs with the IRAS data removed and compared the results
to fits utilising the IRAS data. We find that the integrated intensity
is, on average, overestimated by a factor of 1.3\,$\pm$\,0.6 when the
IRAS data is absent. An inspection of the fits reveals that when IRAS
data is not present the height of the peak depends most strongly on
the SCUBA 450 and 850\,\micron~data and tends to be
exaggerated. Figure~\ref{fig:sed_example} shows a typical SED. The
solid line is a two-component greybody fit to all points, while the
dashed line represents a fit to the data with the IRAS points
removed. Plots of the greybody fits to all sources are available in
the online version of this work.

The percentage error in the fit adds in quadrature to twice the error
in the distance, leading to a luminosity determined to within a
factor of $\sim$\,2 in most cases.


\section{Observations and Data Reduction}
\subsection{The Mopra mm-wave observatory}
The Australia Telescope National Facility (ATNF) `Mopra' telescope is
a 22-m antenna located 26\,km outside the town of Coonabarabran in New
South Wales, Australia. At an elevation of 850 metres above sea level
and a latitude of 31$^{\circ}$ south, it is ideally placed for
observations of the southern Galactic plane.

The receiver is comprised of a set of cryogenically cooled low-noise
SIS mixers, capable of operating at frequencies between 85 and
115\,GHz. A polarisation splitter divides the beam into two orthogonal
linear polarisations (A and B), which are processed separately. The
system is usually configured for dual-polarisation observations, with
both polarisations tuned to the same sky frequency, however,
polarisation B can be independently tuned to a dedicated `pointing'
frequency at the 86.243\,GHz masing transition of SiO. A digital
auto-correlator forms the back-end, providing each polarisation with
available bandwidths ranging from 4 to 256\,MHz, which can be split
into a maximum of 4096 channels.

The antenna is a Cassegrain design  with a 22-m parabolic surface
shaped to maximise the forward gain. In a study performed by
\citet{Ladd2005} the half-power main-beam size was found to vary
between 36\,$\pm$\,3\,\arcsec~and 33\,$\pm$\,2\,\arcsec~over the
frequency range 85\,--\,115\,GHz. The main-beam efficiency varied from
0.49 $\pm$ 0.03 to 0.42 $\pm$ 0.02  over the same range. Sources with
angular extent $<$\,80\,\arcsec~couple well to the Gaussian main beam;
however, the first error lobe, which extends from diameters of
80\,\arcsec~to 160\,\arcsec~contains $\sim$\,1/3 of the power present
in the main beam and a separate set of efficiencies must be used for
calibration of extended sources. We have assumed all sources just fill
the main beam and have corrected all of our data onto the main beam
temperature scale.

For single pointing surveys the only mode of observation offered is
position switching, where the telescope alternates between the science
object and an emission-free sky position. 

Further technical details on the design of the antenna and receiver
systems can be found in the Special Edition of the Journal for
Electrical and Electronic Engineering, Australia, Vol 12, No 2, 1992.


\subsection{The MSX satellite}
The MSX satellite was launched in 1995 by the US Ballistic Missile
Defence Organisation with the principal aim of performing a census of
the mid-infrared sky. The telescope has a 35-cm clear aperture feeding
the SPIRIT-III infrared camera, which has a spatial resolution of
18.3\,\arcsec. Four wavebands dubbed  A, C, D and E are useful for
astronomical research and lie at 8, 12, 14 and 21\,\micron,
respectively. Band A (6.8--10.8\,\micron) is the most sensitive of the
four filters but is contaminated by emission from the strong silicate
band at 9.7\,\micron~and the polycyclic aromatic hydrocarbon (PAH)
band at 8.6\,\micron. Band E (17.2--27.1\,\micron) is less sensitive
by a factor of $\sim$\,20 than A, and is directly analogous to the
broad IRAS 25\,\micron~filter. The instrumentation and survey is
described in detail by \citet*{EganPrice1996}. Calibrated images of
the galactic plane on a common grid are freely available from the
on-line MSX image server at the IPAC archive website at:
http://irsa.ipac.caltech.edu/.


\subsection{Observations}
We observed our sample using the 92\,GHz J\,=\,5\,--\,4 and 110\,GHz
J\,=\,6\,--\,5 transitions of \chthreecn~as well as the 89 and 86\,GHz
J\,=\,1\,--\,0 transitions of \hcop~and \hthirteencop.

The observations reported here were performed as part of a larger
survey using the Mopra antenna, conducted over 5 years, from
2000\,--\,2004, and between the months May\,--\,September. We observed
primarily in the dual-polarisation mode. Each polarisation spanned an
identical bandwidth of 64\,MHz split into 1024 channels, resulting in
a velocity resolution of $\sim$\,0.2\,\kms~over a range of
$\sim$\,180\,\kms. Typical system temperatures ranged from 150\,K to
350\,K over the course of the observations. Position switching was
employed to subtract the sky contribution and bandpass, with
emission-free reference positions chosen at declination offsets
greater than 10\,\arcmin. In the few cases where evidence of emission
was seen in the reference position, new positions were selected
greater than 1\,\degree~away from the Galactic plane. The pointing
accuracy was checked at hourly intervals by observing bright SiO
masers of known position and is estimated to be better than
10\,\arcsec. The pointings towards five sources were consistently off
by between 5 and 15\,\arcsec, due to an error in the observing
schedule file. G5.89$-$0.39 and G19.70$-$0.27 are offset by 5\,\arcsec,
G29.98$-$0.04 is offset by 10\,\arcsec~and G29.96$-$0.02 is offset by
15\,\arcsec. The measured brightness temperature will be diminished by
an unknown factor in these sources.

A single-load chopper wheel was used to calibrate the data onto a
T${\rm _A^{\ast}}$ scale (see \citealt*{KutnerUlich1981}). At Mopra
this was implemented by means of a single ambient temperature
blackbody paddle placed in front of the beam every few minutes. The
data were further calibrated onto a main beam temperature scale
(T$_{\rm MB}$) by dividing by the main beam efficiencies quoted in
\citet{Ladd2005}. 

The typical RMS noise on the T$_{\rm MB}$ calibrated \chthreecn~spectra
is 80\,mK for integrations on-source of approximately 30
minutes. Integration times of 5 and 15 minutes lead to a RMS noise of
$\sim$\,200\,mK and $\sim$\,100\,mK in the \hcop~and
\hthirteencop~spectra, respectively.


\subsection{Data Reduction}
\subsubsection{Mopra spectral line data}
Raw data in {\it rpfits} format were initially reduced using the
  SPC\,\footnote{http://www.atnf.csiro.au/software/}$^,$\,\footnote{A
  tcl/tk user interface to SPC known as Data From Mopra (DFM) is
  available from http://www.phys.unsw.edu.au/mopra/software/} package
  developed by the ATNF and custom routines used to correct for errors
  in the headers\,\footnote{During 2000\,--\,2003 Mopra suffered from
  a rounding error in the receiver control software which lead to an
  incorrect velocity scale in the data. A pre-processing script for
  correcting the data is available from the ATNF Mopra website:
  http://www.mopra.atnf.csiro.au}. During the reduction, the spectra
  were sky-subtracted and the two polarisations averaged together
  before a low order polynomial was subtracted from the baseline. The
  reduced data were then read in to the XS\,\footnote{XS is written by
  Per   Bergman at the Onsala Space Observatory and is available via
  FTP on request.} package where higher order polynomials or sine
  waves were subtracted if necessary. Finally all the data were
  converted into the CLASS\,\footnote{CLASS is part of the GILDAS data
  reduction package available at http://www.iram.fr/IRAMFR/GILDAS/}
  format and the spectra were divided by the  year-to-year efficiency,
  ${\rm \eta_{yr}}$, before calibrating onto the main beam brightness
  temperature (${\rm T_{\rm MB}}$) scale.


\subsubsection{Mid-infrared data}
Individual 8 and 21\micron~images centred on the methanol maser
position were acquired from the MSX image server. The MSX images have
a stated positional accuracy of 10\,\arcsec. In an effort to improve
registration between datasets, we have utilised the 2MASS near-IR
images, which have a pointing accuracy generally better than
1\,\arcsec. If a constant offset was observed between two or more
bright sources present in both data sets, we shifted the MSX
coordinates to match those of the the 2MASS images. We considered a
maser site or UCH{\scriptsize II} region to be associated with a MSX
source if it fell within 9$''$ (half of the full-width at half-maximum
(FWHM) of the MSX point spread function) of the peak of the infrared
source. The MSX fields often suffered from confusion and blending,
especially at 8\,\micron. Where possible, we measured the integrated
intensity by drawing a polygon around the source and summing the pixel
values. A pixel-averaged sky value was measured from an emission-free
region and subtracted from the resulting figure to find the net
integrated flux in units of ${\rm W m^{-2}\,sr^{-1}}$. Conversion to
Janskys ({$\rm W\,m^{-2}\,Hz^{-1}\times 10^{26}$}) was achieved by
dividing by the bandwidth of the filter, and multiplying by the
6\,$\times$\,6\arcsec~pixel area in steradians (8.4615$\times
10^{-10}$ Sr). An additional multiplicative factor of 1.113 was
required to correct for the Gaussian response of the individual pixels
\citep{EganPrice1996}. The final scaling factors to convert fluxes in
${\rm W\,m^{-2}\,sr^{-1}}$ to Jy were: 6.84$\times$10$^3$,
2.74$\times$10$^4$, 3.08$\times$10$^4$ and 2.37$\times$10$^4$ for the
four bands A, C, D and E respectively. On inspection of the
8\,\micron~images we suspect much of the extended emission seen is due
PAHs. If a clear distinction was evident between the source and the
extended emission, we measured the sky value from average level of the
extended emission close to the source. 


\section {Results}
\subsection{Spectral line detections}
We present a summary of the lines detected towards our sources in
Table~\ref{tab:detections}. \chthreecn\,(5\,--\,4) emission was detected
towards 58 sources (70\%), 43 of which are new detections. We followed
up the \chthreecn\,(5\,--\,4) detections in 33 sources by searching for
the 110\,GHz \chthreecn\,(6\,--\,5) transition and detected the line
in 24. \hcop~was
detected towards 82 sources (99\%), and \hthirteencop~was detected
towards 80 sources (98\%). Detected spectral lines peak at least
2-$\sigma$ above the baseline and in most cases have a signal to noise
ratio greater than four. 

Figure~\ref{fig:g0.55} presents example spectra for the bright source
G0.55+0.85 alongside the 21\,\micron~MSX image. Annotated on the MSX
image are the positions of the methanol masers (crosses) and the
36\,\arcsec~FWHM size of the Mopra beam at 86\,GHz (circle). Squares mark
the peaks of the compact 8 GHz continuum emission detected by
\citet{Walsh1998}. Spectra and MSX images for all sources are
available as additional online material. 
\begin{figure}
  \begin{center}
    \includegraphics[width=7.9cm, trim=0 -0 -0 0]{figs/fig_2.epsi}
    \caption{~\chthreecn, \hcop~and \hthirteencop~spectra for the bright
      source G0.55$+$0.85 alongside the 21\micron~MSX image (grey-scale \&
      contours). On the image squares mark known UCH{\scriptsize II}
      regions and crosses mark 6.7\,GHz methanol-maser sites. The
      circle shows the 36\,\arcsec~Mopra beam, centred on the maser
      position. The \chthreecn~K-components are well fit by Gaussians,
      except for the K\,=\,4 line, which may be affected by baseline
      instabilities. The dashed line marks the V${\rm _{LSR}}$, taken
      from a Gaussian fit to the \hthirteencop~line. The \hcop~line
      profile is asymmetric and blue-shifted with respect to the
      \hthirteencop~line, possibly indicating in-falling gas motions.} 
    \label{fig:g0.55}
  \end{center}
\end{figure}

Individual components within the \chthreecn~K-ladder were fit
simultaneously with multiple Gaussians. The separations were fixed to
their theoretical values and the linewidths constrained to have the
same value in the fit. Table~\ref{tab:mclineparams} lists the Gaussian
parameters of the fits to both the \chthreecn\,(5\,--\,4) and
(6\,--\,5) lines. Column one is the Galactic source name, columns two
and three are the V${\rm _{LSR}}$ and FWHM, columns four to eight
contain the fluxes, in K\,\kms, of the K\,=\,0 to K\,=\,4 rotational
ladder components. Quoted 1\,$\sigma$ uncertainties in the fits are
those reported by the CLASS analysis package used to fit the data. The
1\,$\sigma$ noise on the spectra of sources without
detected \chthreecn\,(5--4) are given in Table~\ref{tab:ch3cn_nond}.

Absolute calibration errors were estimated from multiple observations
of individual sources and are generally on the order of 30\,\%. The
low signal to noise ratios in many spectra made identification of
asymmetries or line wings difficult and we assume the individual K
lines were well fit by Gaussians. During our analysis we found that
the K\,=\,4 component of the \chthreecn\,(5--4) spectrum was almost
always anomalously strong and sometimes appeared shifted lower in
frequency with respect to the other components. It is possible we are
beginning to see excitation by higher temperature gas within the hot
core, however, the signal to noise ratio for almost all the K\,=\,4
detections is very low and in many cases broad features are seen which
have no counterpart in the other K-components, indicating possible
baseline problems.

\begin{figure}
  \begin{center}
    \includegraphics[height=7.9cm, angle=-90, trim=0 -0 -0 0]{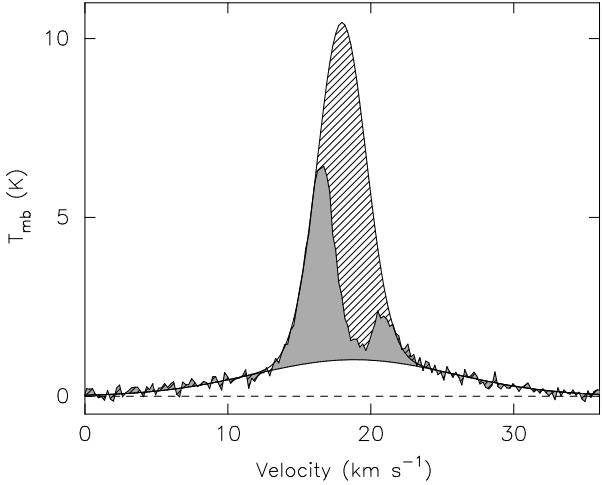}
    \caption{~An illustration showing the different methods of
      measuring the self-absorbed \hcop~line profiles. We measure the
      line intensity by integrating under the profile (grey area) and by
      fitting a Gaussian, constrained by the sides of the line and the
      velocity of the optically thin \hthirteencop~(hatched + grey
      area). The contribution from high velocity outflow-wings are
      omitted from both measurements.} 
    \label{fig:fit_cartoon}
  \end{center}
\end{figure}

In most cases the detected \hthirteencop~lines are well fit with a
single Gaussian line, show little evidence of self absorption or line
wings, and are assumed to be optically thin. In contrast, the
\hcop~profiles are complex, often exhibiting two or more components
(e.g. G8.68$-$0.37). These profiles may be interpreted as either
multiple clouds along the line of sight or a single broad emission
line with a self-absorption dip. In an attempt to distinguish between
these two cases we examined the profiles for evidence of multiple
components. Where a single \hthirteencop~ component was visible at the
velocity of the dip in the \hcop~line we attempted to fit the
double-peaked profile with a single Gaussian, based on the assumption
that the line was strongly self-absorbed. In practise, we blanked the
absorption dip from the fitting routine so that the Gaussian fit was
constrained only by the sides of the line. At best this gives an upper
limit to the intensity of the line. As a lower limit we measured the
integrated intensity under the line-profile, after subtracting
Gaussian fits to any line wings. Figure~\ref{fig:fit_cartoon}
illustrates this method using the extreme case of G16.86$-$2.16. The
results of these measurements are reported in Table~6.

Twenty-six sources detected in \hcop~have residual line wings distinct
from the main line, pointing to the possible existence of molecular
outflows. We fitted and subtracted detected line wings with a single
broad Gaussian {\it before} attempting to analyse the main
line. Sixteen sources are found to contain blends of two or more
\hcop~lines, which we attempted to fit with individual Gaussians in
order to separate the main line from the blend. Intensities quoted in
Table~6 are for the main line only and do {\it not} contain
contributions from line wings or blended lines. The Gaussian fit
parameters to the line wings and blended lines are recorded separately
in Table~\ref{tab:hcop_linewings}.


\subsection{Mid-infrared associations}
We present the association between the maser-selected sample and
mid-infrared MSX sources in Table~\ref{tab:detections}. Sixty-eight
positions observed have mid-IR emission peaking within 10$''$. Of
these, 3 are detected at 8\,\micron~but not at 21\,\micron, and 7 are
detected at 21\,\micron~but not at 8\,\micron. 15 positions show no
evidence of a mid-infrared emission, however 12 of these are
identified as MSX `dark-clouds', seen in absorption against the
Galactic plane at 8\,\micron. A single source, G30.79$+$0.20, is
detected as a point-source at 21\,\micron~but is embedded within a
dark cloud at 8\,\micron. 

Of the 80 sources detected in emission or absorption in the mid-IR, 57
have `compact' or unresolved morphologies, 17 are extended and the
remainder are confused.


\section{Derived Physical Parameters}
\subsection{Rotational Diagram Analysis}
To estimate the temperature and column density of \chthreecn, we have
used the rotational diagram (RD) analysis first introduced by
\citet*{Hollis1982}. The method has been developed and expanded on by
a number of authors (\citealt*{LorenMundy1984}, \citealt*{Turner1991},
\citealt*{GoldsmithLanger1999}). We provide a brief summary here for
convenience.

By solving the radiative transfer equation, assuming low optical depth
(i.e. $\tau \ll 1$), we arrive at the well known relation between upper
state column density, ${\rm N_u}$, and measured line intensity, ${\rm
  \int_{-\infty}^{\infty}T_b\,dv}$
\begin{equation}\label{eq:column_density}
  {\rm N_u = \frac{8\pi k \nu_{ul}^2}{hc^3A_{ul}}\int_{-\infty}^{\infty}T_b\,dv},
\end{equation}
where ${\rm \nu_{ul}}$ is the frequency of the transition in Hz and
${\rm A_{ul}}$ is the Einstein A coefficient in s$^{-1}$. Line
intensity is in units of K\,\kms. All other constants take their usual
values in SI units. Assuming local thermal equilibrium (LTE) (i.e. the
energy levels are populated according to a Boltzmann distribution
characterised by a single temperature, T), the upper state column
density is related to the total column density, ${\rm N}$, by
\begin{equation}\label{eq:boltzmann_relation}
  {\rm N_u=\frac{N g_u}{Q(T)} e^{-E_u/k T}},
\end{equation}
where ${\rm g_u}$ is the degeneracy of the upper state, ${\rm E_u}$ is
the energy of the upper state in joules and ${\rm Q(T)}$ is the
partition function. Rearranging and taking the natural log of both
sides we find 
\begin{equation}
  {\rm \ln\left(\frac{N_u}{g_u}\right)=\ln\left(\frac{N}{Q(T)}\right)-\frac{E_u}{kT}}.
\end{equation}
A straight line fitted to a plot of ${\rm \ln(N_u/g_u)}$ versus ${\rm
E_u/k}$ will have a slope of ${\rm 1/T}$ and an intercept of ${\rm
\ln(N/Q(T))}$. Temperatures found from this method are referred to as
rotational temperatures (T$_{\rm rot}$) in the literature. In our
analysis we used the partition function for \chthreecn~quoted in
\cite{Araya2005}
\begin{equation}
{\rm Q(T) = \frac{3.89T_{rot}^{1.5}}{(1-e^{-524.8/T_{rot}})^2}}.
\end{equation}

Inherent in the RD method are some formal assumptions which may effect
the interpretation of the results. Firstly, the emitting region is
assumed to be optically thin. If optical depths are high the measured
line intensities will not reflect the column densities of the
levels. Optical depth effects will be evident in the rotational
diagram as deviations of the intensities from a straight line and a
flattening of the slope \citep{GoldsmithLanger1999}, leading to
anomalously large values for T$_{\rm rot}$. 

We have attempted to iteratively correct individual ${\rm N_u/g_u}$
values by muliplying by the optical depth correction factor, ${\rm
  C_{\tau}=\tau   /(1-e^{-\tau})}$, after the method of
\citet{Remijan2004}. Assuming Gaussian line profiles, the line-centre
optical depth is given by  
\begin{equation}\label{eq:tau_column}
  {\rm \tau = \frac{c^3\sqrt{4\,\ln\,2}}{8\pi \nu^3\sqrt{\pi}\Delta v}N_u A_{ul}\left(e^{h\nu/kT_{rot}}-1\right)},
\end{equation}
where ${\rm \Delta v}$ is the FWHM of the line profile in \kms~(see
\citet{GoldsmithLanger1999} and \citet{Remijan2004} for a formal
description). In our calculations we initially assumed the source
fills the beam. We find that the corrected optical depth is always
$\ll$\,1, even in the K=0 transitions, and the correction factors are
less than 1\,\%. The FWHM Mopra beam at 92 GHz is
$\sim$\,36\,\arcsec. If previous high-resolution surveys
(e.g. \citealt{Remijan2004}, \citealt*{Wilner1994}) are indicative,
the angular size of \chthreecn~emission will be $<$\,10\,\arcsec. Our
measurements of \chthreecn~intensity are thus likely to be affected by
a dilution factor of $>$\,13 and the measured optical depth, and
therefore the correction factors, are anomalously low because of this.

If we then assume a 7\,\arcsec~source size and determine the correction
factor, we find that the rotational temperature is corrected down by
$\sim$10\,\% and the total column density is corrected up by
$\sim$\,6\%, on average. In their analysis of Sagittarius B2 
\citet{Nummelin2000} chose a source size
which minimises the $\chi^2$ value on their RD fits. They find
that for some molecules the best fit to the data is obtained for beam
filling factors $\ll$\,1. We find that in most cases the quality of
our fits {\it deteriorate} as the beam filling factor is decreased,
most likely because our fits our not well constrained due to low
signal\,/\,noise data. Thus we use a beam filling factor of 1 in our
final calculations, and beam-averaged column densities stemming from
this analysis should be considered a lower limit.

Secondly, the emitting region is assumed to be in LTE where the
population within a K ladder is characterised by a single excitation
temperature, described by a Boltzmann distribution. For the non-LTE
case the temperature and column density will be underestimated.

\begin{figure}
  \begin{center}
    \includegraphics[height=7.9cm, angle=-90, trim=0 -0 -0 0]{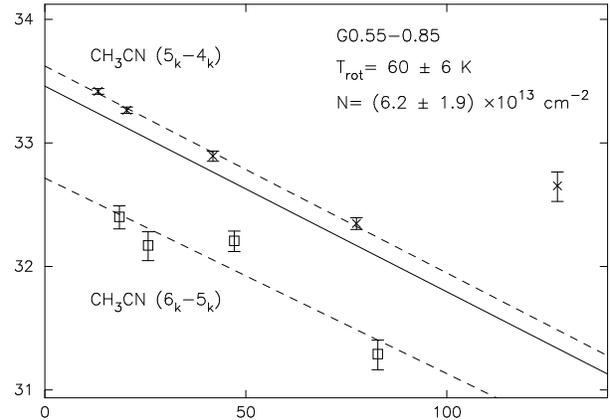}
    \caption{~\chthreecn~rotational diagram of ${\rm \ln (N_u/g_u)}$ vs
      ${\rm E_u/k}$ for the source G0.55$-$0.85. The J\,=\,5\,--\,4
      K-components are plotted with crosses and the J\,=\,6\,--\,5
      K-components are plotted with squares. Error bars denote the
      1\,$\sigma$ errors in the Gaussian fit to individual lines. The
      rotational temperature, ${\rm T_{rot}}$, and total column density,
      N, may be found from the slope and y-axis intercept of a straight
      line fit to the data points. We fit the K-ladders separately
      (dashed lines) and in most cases took the weighted mean of the
      results as our final values for ${\rm T_{rot}}$ and N (solid
      line). The error in ${\rm T_{rot}}$ is dependant only on the
      relative calibration between lines in a single spectrum. The error
      in N is dependant on the absolute calibration, and is
      approximately 30\,\%.} 
    \label{fig:rotdiag_example}
  \end{center}
\end{figure}

Figure~\ref{fig:rotdiag_example} shows a typical example of a
rotational diagram made using the J\,=\,6\,--\,5 and J\,=\,5\,--\,4
transitions. Calibration errors between the J transitions are
approximately 30\,\% and manifest themselves as a vertical offset
between the two sets of points in the graph. The change in the FWHM of
the Mopra beam between 92 and 110\,GHz has been measured by
\citet{Ladd2005}, and the intensities of the J\,=\,6\,--\,5 lines
have been divided by the relative beam filling factor, given by
${\rm (\theta_{110\,GHz}/\theta_{93\,GHz})^2}$ =
(34\,\arcsec/36\,\arcsec)$^2$. As the K components within a single J 
transition were observed simultaneously, the relative errors are
smaller and arise from the quality of the detection and stability of
the bandpass. Error bars mark the 1\,$\sigma$ error in the column
density of individual K components, as determined by Gaussian
fitting. In the analysis we fit the J\,=\,6\,--\,5 and J\,=\,5\,--\,4
data separately, shown in Figure~\ref{fig:rotdiag_example} as dashed
lines. We took as our final result the weighted average of the derived
T${\rm _{rot}}$ and N, plotted on the diagram as a solid line. The
rotational diagrams for the remaining sources are available in the
online edition of this paper. 

The results of our rotational diagram analysis are presented in
Table~\ref{tab:trot}. In general, the observed intensities were well
fit by a straight line, except for G10.47+0.03 which exhibited
anomalous line ratios and for which a meaningful temperature could not
be derived. Rotational temperatures range from 28\,K to 166\,K with an
average value of 57\,K. Temperatures derived from sources where we
detected less than four \chthreecn~K components are not well
constrained, but are still useful for comparative purposes.

We compared distributions of T${\rm _{rot}}$ for UHC{\scriptsize II}
regions and isolated masers, but a KS-test did not find any
significant difference in rotational temperature between them.

It is interesting to compare our derived rotational temperatures and
column densities to previously published results. Eleven sources reported
here overlap with previous studies by \citet{Pankonin2001},
\citet{Hatchell1998}, \citet*{Olmi1993} and
\citet*{Churchwell1992}. Table~\ref{tab:comparison} is a reproduction
of Table~6 from \citet{Pankonin2001}, expanded to include this
paper. In it we compare values of derived rotational temperature and
column density. Our values for T$_{\rm rot}$ are generally lower,
typically by a factor of $\sim$\,2, except when compared to values
derived by \citet{Olmi1993}, who also employ the \chthreecn~(5--4)
transition in their analysis. The other studies have included data
from the J\,=\,12\,--\,11 transitions or higher. The J\,=\,5\,--\,4
and J\,=\,6\,--\,5 transitions have upper-state energies spanning
$\sim$\,8\,--\,130\,K compared to $\sim$\,68\,--\,500\,K for the
J\,=\,12\,--\,11 transition. It is likely our survey is sensitive to
cooler gas, as evidenced by our lower derived temperatures, whereas
the surveys utilising higher energy transitions in their analysis are
sensitive to hotter gas. These assertions are supported by recent
observations by \citet{Araya2005} who found it necessary to fit two
temperatures to their observations of the J\,=\,5\,--\,4, 6\,--\,5,
8\,--\,7 and 12\,--\,11 transitions of \chthreecn. The lower energy
lines yield typical temperatures of 35\,K whereas the higher energy
lines are well fit by temperatures above 90\,K, suggesting the
existence of a hot molecular clump embedded in a cooler envelope, or
simply a temperature gradient. Further evidence of a temperature
gradient in hot cores is presented by \citet{Olmi2003}, who observed
\chthreecn\,(6\,--\,5) in G29.96$-$0.02. An analysis of the angular
diameter of the emission measured in each K component shows that
higher energy transitions are excited closer to the centre of the hot
core.

Column densities derived here are comparable to those reported in
other work. They are higher by factors of $\sim$\,1.5\,--\,2 compared to
the work of \citet{Pankonin2001}, but significantly lower than earlier
surveys by \citet*{Olmi1993} and \citet*{Churchwell1992}. Caution is
needed, however, in making a direct comparison of column-densities
between surveys, as the beam-sizes and excitation conditions of
different observations are not well matched. There is also confusion
in the literature with regard to the correct partition function for
\chthreecn, which may lead to the column-densities being miscalculated
by as much as a factor of 5 \citep{Araya2005}. The high figures
derived in this paper compared to \citet{Pankonin2001} probably
reflect the greater volume of lower temperature gas probed by the
36\,\arcsec~Mopra beam compared to more compact, warmer gas probed by
other surveys.


\subsection{\hcop~Column Density and Abundance}
We derive \hcop~column densities assuming optically thick \hcop~and
optically thin \hthirteencop~lines. 

The expression for the measured brightness temperature, ${\rm T_b}$,
of a molecular transition is:
\begin{equation}\label{eq:measured_brightness}
  {\rm \frac{T_b}{\eta_{bf}}= T_r =[J_{\nu}(T_s)-J_{\nu}(T_{bg})](1-e^{-\tau_{\nu} })},
\end{equation}
where 
\begin{equation}\label{planck_brightness}
  {\rm J_{\nu}(T) = \frac{h\nu_u}{k}\frac{1}{(e^{h\nu_u/kT}-1)}},
\end{equation}
and where ${\rm T_s}$ is the temperature of the source, ${\rm T_{bg}}$
is the temperature of the background radiation (2.73\,K), $\tau$ is
peak optical depth, and ${\rm \eta_{bf}}$ is the beam filling factor,
given by $\Omega_s / \Omega_b$. Assuming the \hcop~line is optically
thick, ($1-e^{-\tau}\approx 1$), we can calculate an excitation
temperature according to
\begin{equation}
  {\rm T_{ex} = T_s = \frac{h \nu_u}{k}\left[\ln\left(1+\frac{(h\nu_u/k)}{T_{thick}+J_{\nu}(T_{bg})}  \right) \right]^{-1}},
\end{equation}
where ${\rm T_{thick}= T_{r}}$ is the beam corrected brightness
temperature of the \hcop~line. We assume the \hcop~and
\hthirteencop~emission arises from the same gas and shares a common
excitation temperature. Given a known excitation temperature, the
optical depth of the \hthirteencop~line may be found from
\begin{equation}
  {\rm \tau_{thin} = - \ln\left[1-\frac{T_{thin}}{[T_{ex}-J_{\nu}(T_{bg})]}\right]},
\end{equation}
where ${\rm T_{thin}=T_r}$ is the beam corrected brightness
temperature of the \hthirteencop~line. 

Equations~\ref{eq:boltzmann_relation} and~\ref{eq:tau_column} may then
be applied to find the total \hthirteencop~column column density,
again assuming LTE conditions. The partition function for linear
rotors is well approximated by ${\rm Q(T) = (k T_{ex})/(h B)}$, where
B is the rotational constant in Hz (43377\,MHz for \hthirteencop). The
\hcop~and \hthirteencop~column densities are related through their
relative abundance ratio, 
\begin{equation}
  {\rm X = \frac{[\hcop]}{[\hthirteencop]} = \frac {N_{HCO^+}}{ N_{H^{13}CO^+}}},
\end{equation}
which is partially dependant on the fraction of $^{\rm 13}$C present
in the interstellar medium. A Galactic gradient in the $^{\rm
12}$C\,/\,$^{\rm 13}$C ratio, ranging from $\sim$\,20 to $\sim$\,70,
has been measured using several methods (see \citet{Savage2002} and
references therein). The measurement uncertainties are large, however,
and the ratio of [\hcop]\,/\,[\hthirteencop] may also be affected by
chemistry. Thus, we have assumed a constant X\,=\,50 for all our
calculations.

A major source of error in this analysis stems from the assumption
that the \hcop~lines provide a meaningful estimate of the excitation
temperature. In many cases the line profiles show evidence of self
absorption and are poorly fit by simple Gaussians. As outlined in
\S~3.1 we used two methods to estimate the \hcop~brightness
temperature. In the first method we measure the brightness temperature
from a Gaussian fit to the profile, by masking off the absorption
dip. In the second method we took as the value for brightness
temperature the highest peak in the profile (see
Figure~\ref{fig:fit_cartoon}). If the emission is smaller than the
beam, the 
filling factor, ${\rm \eta_{bf}}$, will be less than one and the
excitation temperature will be underestimated. Using the above methods
we find average excitation temperatures of $\sim$\,7.5\,K and
$\sim$\,6.1\,K, respectively. These values are considerably lower
than the $\sim$\,15\,K temperatures quoted in previous work
(e.g. \citealt{Girart2000}) and we conclude that the emission is beam
diluted in a significant fraction of our observations. In our final
analysis we have assumed an excitation temperature of 15\,K and the
column densities are calculated solely from the
\hthirteencop~lines. The \hthirteencop~lines were well fit with
Gaussians and are consistent with being optically thin in most
cases. The results of the analysis are presented in
Table~\ref{tab:hcop_deriv}. As the source size is unknown the values
quoted are beam-averaged, assuming a filling factor of one.


\section{Analysis \& Discussion}
We detect \chthreecn~towards all classes of sources: UCH{\scriptsize}
II regions, isolated masers and maserless cores. This result confirms
a direct link between each of these objects and massive star
formation. In the following analysis we will discuss the implications
for a possible evolutionary sequence.


\subsection{The presence of an UCH{\scriptsize II} region}
The \chthreecn~detection rate towards maser-sources associated with an
UCH{\scriptsize II} region (radio-loud), compared to isolated maser
sources (radio-quiet) is strikingly different. 
\begin{figure*}
  \begin{center}
    \includegraphics[height=6.9cm, trim=0 -0 -0 0]{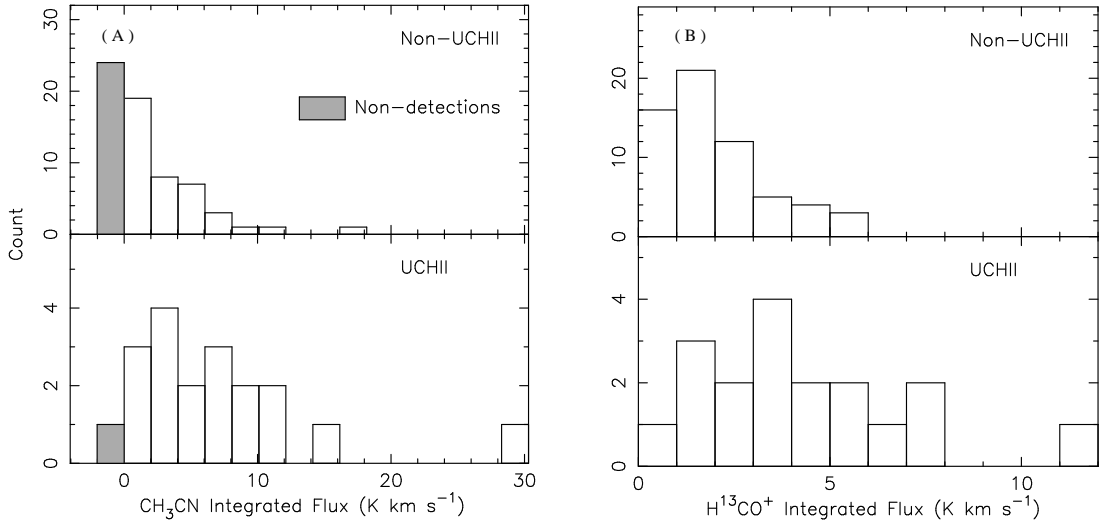}
    \caption{~{\bf  A}) Distributions of integrated
      \chthreecn\,(5\,--\,4) flux, summed over all detected
      K-components, for sources with UCH{\scriptsize II} regions
      within the Mopra beam (bottom) and those without (top). The
      shaded bar represents the number of \chthreecn~non-detections
      for each population. A KS-test yields a maximum difference of
      0.48 between the distributions, with a probability of 0.13\,\%
      of being drawn from the same parent distribution. {\bf B})
      Distributions of integrated \hthirteencop\,(1\,--\,0) flux for
      the same populations. Lines associated with UCH{\scriptsize II}
      regions are markedly brighter than those without. A KS-test
      yields a difference of 0.55 between the distributions, with an
      associated probability of 0.02\,\% of being drawn from the same
      parent distribution. A similar result has been found for the
      \hcop~lines.} 
    \label{fig:ch3cna_uchii}
  \end{center}
\end{figure*}

Figure~\ref{fig:ch3cna_uchii}-A shows the distributions of
integrated \chthreecn\,(5\,--\,4) intensity for the two  classes of
source, including non-detections, which are summed in the shaded
bar. \chthreecn~was detected towards 18 out of the 19 radio-loud
sources (95\,\%) compared to only 40 out of 64 radio-quiet sources
(63\,\%). Furthermore, the average integrated \chthreecn\,(5\,--\,4)
intensity measured towards the radio-loud sources is 7.4\,K\,\kms,
stronger than the average measured towards the radio-quiet sources,
which is 3.5 K\,\kms. We use a Kolomogorov-Smirnov (KS) test to
measure the difference between the distributions of the two
classes. Non-detections are included as upper limits, measured from
the RMS noise in the spectra, and by assuming K\,=\,0 and K\,=\,1
\chthreecn~components of linewidth 4\,\kms~are present at a
3\,$\sigma$ level. Using this conservative estimate a KS-test
indicates there is only a 0.13\,\% probability of the samples being
drawn from the same population.  Probabilities below $\sim$\,1\,\% 
indicate that the distributions are significantly different.

A similar result applies for
\hthirteencop. Figure~\ref{fig:ch3cna_uchii}-B shows the distributions
of \hthirteencop~intensity for the two classes of objects. The mean
intensity for the radio-loud sources is 4.3 K\,\kms~compared to 1.9
K\,\kms~for the radio-quiet sources. A KS-test returns a probability
of 0.02\,\% that the distributions are similar.  

We find the \hcop~lines are also brighter towards radio-loud sources,
however we do not consider them in this analysis as they show evidence
of optical depth effects which may confuse their interpretation.  

\subsubsection{The effect of distance on the results}
\begin{figure}
  \begin{center}
    \includegraphics[height=6.9cm, angle=-90, trim=0 -0 -0 0]{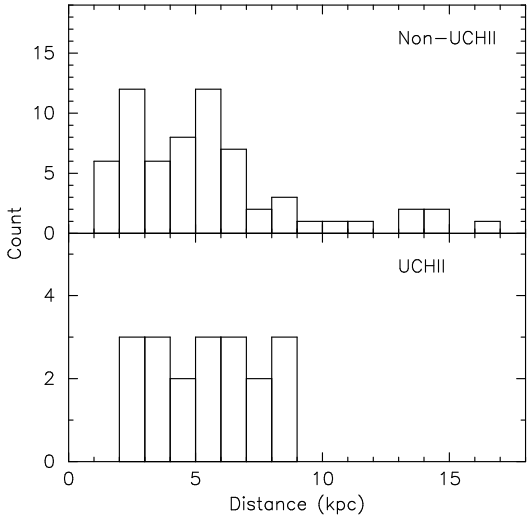}
    \caption{~Distributions of distances in kiloparsecs for sources
      with and without associated UCH{\scriptsize II} regions. A KS-test
      yields a probability of 53\,\% that samples are consistent with
      being drawn from the same population and thus probe the same
      distances.} 
    \label{fig:hist_dist_uchii}
  \end{center}
\end{figure}

A higher \chthreecn~detection rate towards UCH{\scriptsize II} regions
is consistent with a fall-off in flux with distance in the original
radio survey. Our classification of sources into radio-loud and
radio-quiet regions may be artificial, and undetected UCH{\scriptsize
  II} regions may simply be a result of the sources being further
away. Closer examination shows this not to be the case. The
distribution of distances to the two populations is shown in
Figure~\ref{fig:hist_dist_uchii}. No UCH{\scriptsize II} regions have
been detected further than 9\,kpc away, however only 9 out of 83
sources lie at greater distances. A comparison yields an 53\,\%
probability that the classes are consistent with being drawn from the
same population and thus probe the same distance range.

\begin{figure}
  \begin{center}
    \includegraphics[height=7.9cm, angle=-90, trim=0 -0 -0 0]{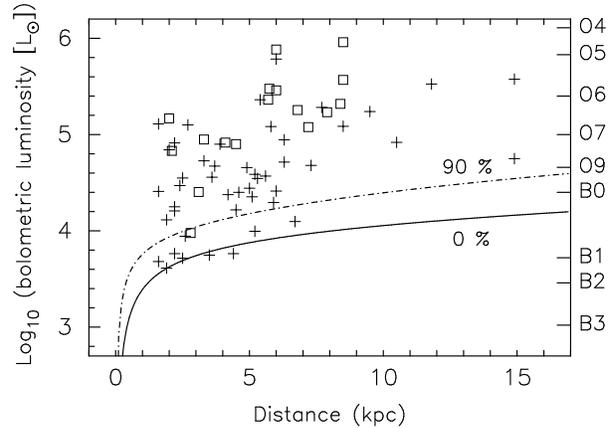}
    \caption{~A plot of distance versus bolometric luminosity for the
      sample. The solid line shows the 3\,$\sigma$ detection limit for
      8.67\,GHz radio continuum emission, assuming all the Lyman-continuum
      flux contributes to creating an UCH{\scriptsize II} region. The
      dashed line shows how the detection limit changes if 90\,\% of the
      ionising photons are absorbed by dust. UCH{\scriptsize II} regions
      are represented by squares, and isolated maser sources by crosses.} 
    \label{fig:xy_dist_lum}
  \end{center}
\end{figure}
\begin{figure*}
  \begin{center}
    \includegraphics[height=6.9cm, trim=0 -0 -0 0]{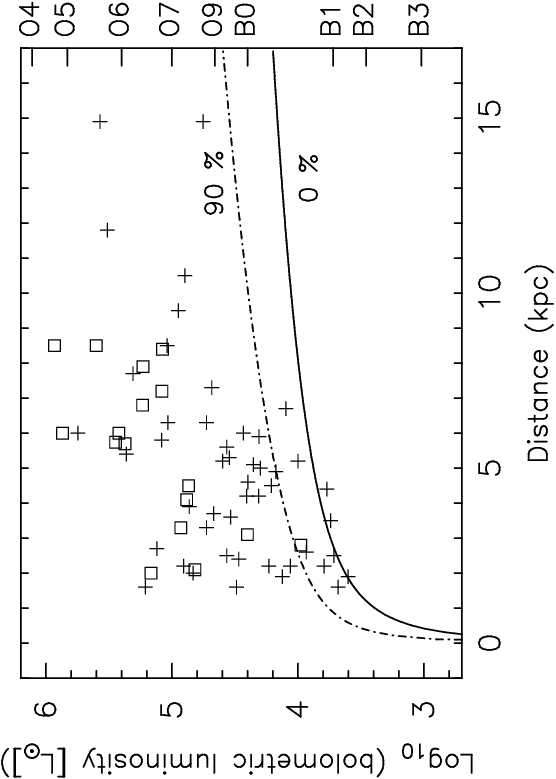}
    \caption{~As for Figure \ref{fig:ch3cna_uchii} but limited to
    sources which are luminous enough to have detectable
    UCH{\scriptsize II} regions, assuming no less than 90\,\% of the
    ionising photons are attenuated. {\bf A}) A KS-test yields a
    maximum difference of 0.50 between the \chthreecn~distributions,
    with an associated probability of 0.93\,\% of being drawn from the
    same parent distribution. {\bf B}) Similarly the maximum
    difference between the \hthirteencop~populations is 0.56, with an
    associated probability of 0.12\,\%.} 
    \label{fig:ch3cna_uchii_limit}
  \end{center}
\end{figure*}

We further examine the detection limits on the \citet{Walsh1998} 8.64 GHz radio
survey from which our sample was drawn. The quoted 1\,$\sigma$
sensitivity is $\sim$\,1\,mJy. Assuming a 3\,$\sigma$ detection, the
limiting distance, D, in kpc at 
which free-free radio emission from a UCH{\scriptsize II} region is
detectable, may be calculated from \citet*{KurtzChurchwellWood1994}: 
\begin{equation}
  {\rm \left(\frac{S_{\nu}}{Jy}\right)= 1.32 \times 10^{-49} N_L \left(\frac{D}{kpc}\right)^{-2}
  a(\nu,T_{e})\left(\frac{\nu}{GHz}\right)^{-0.1} \left(\frac{T_{e}}{K}\right)}
\end{equation}
where ${\rm \nu}$ = 8.64 GHz is the frequency, ${\rm T_e \approx
10^4}$\,K is the electron temperature, ${\rm a(\nu,T_{e})}$ is a
factor of order unity tabulated by \citet*{Mezger1967} and ${\rm N_L}$
is the number of Lyman-continuum photons. Values of ${\rm N_L}$ for
early-type stars are tabulated by \citet{Panagia1973}, however, in
practise some fraction of the emitted ionising photons will be
absorbed by dust before ionising the surrounding medium. In a study of
UCH{\scriptsize II} regions, \citet{KurtzChurchwellWood1994} found
that between 50\,\% and 90\,\% of the ionising photons are absorbed
for the majority of their sample. Figure~\ref{fig:xy_dist_lum} is a
plot of luminosity against distance, in kpc, for all the sources whose
luminosity could be determined. The solid line marks the 3\,$\sigma$
sensitivity limit to 8.64 GHz radio continuum, assuming no
Lyman-continuum photons are absorbed, while the dotted line shows the
sensitivity limit assuming 90\,\% of all ionizing photons are
absorbed. The luminosities of our sources were determined by fitting a
2-component greybody to available MSX, SCUBA, and IRAS fluxes as
discussed earlier and are accurate to a factor of $\sim$\,2. For those
sources above the 90\,\% absorption limit, UCH{\scriptsize II} regions
would have been detected if they existed, and it is most unlikely that
the comparison between \chthreecn~detected for UCH{\scriptsize II}
regions and isolated maser sources is biased. We
plot Figure~\ref{fig:ch3cna_uchii_limit}-A as per
Figure~\ref{fig:ch3cna_uchii}, but include only sources luminous
enough to have detectable radio-continuum emission, assuming 90\,\%
attenuation of ionising photons. We find that the
\chthreecn~distributions are still significantly different, with only
a 0.9\,\% probability of being drawn from the same population.
The \hthirteencop~intensity distribution for the two classes, filtered
to the same luminosity limit, is shown in
Figure~\ref{fig:ch3cna_uchii_limit}-B. A KS-test returns a 0.1\,\%
probability of them being drawn from the same population. 

The apparent enhancement in {\it line brightness} towards
UCH{\scriptsize II} regions could be mimicked by the effect of beam
dilution. A previous high resolution survey for \chthreecn~has found
the typical size of the emitting region to be $<$\,10$''$
\citep{Remijan2004}. Assuming a constant spatial size, the effect of
beam dilution on the \chthreecn~brightness temperature will depend on
the angular size and hence distance to the source. We have shown that
the radio-loud and radio-quiet samples probe the same distance range
and thus neither class is biased towards being
nearer.

\begin{figure}
  \begin{center}
    \includegraphics[height=7.9cm, angle=-90, trim=0 -0 -0 0]{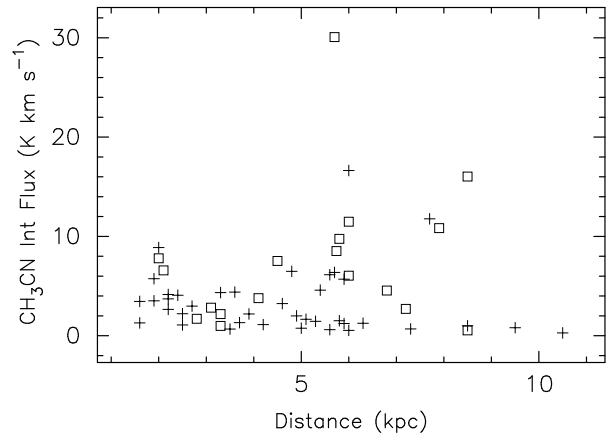}
    \caption{~Plot of integrated \chthreecn~(5\,--\,4) flux vs
      kinematic distance, in kpc. UCH{\scriptsize II} regions are marked
      with squares and isolated maser sources are marked with
      crosses. No correlation is evident.} 
    \label{fig:xy_ch3cn_dist_uchii}
  \end{center}
\end{figure}

Figure~\ref{fig:xy_ch3cn_dist_uchii} plots measured
\chthreecn~(5\,--\,4) flux against kinematic distance and we do  not
find any correlation. As an additional test we can recover the
line-luminosity by multiplying by the square of the distance. This has
the effect of increasing the spread in the brightness distribution for 
both samples, however a KS-test returns a 1.5\,\% probability in the
case of \chthreecn~and a 1.0\,\% probability in the case of
\hthirteencop~that the  radio-quiet and radio loud line-luminosity
distributions are drawn from the same population. The average
line-luminosity remain higher for the radio-loud class of objects. 

\subsubsection{Interpretation.}
Previous work on hot cores suggests \chthreecn~is the tracer of choice
for massive protostars in the hot core phase
(e.g. \citealt{Kurtz2000}), however we find \chthreecn~is
preferentially associated with UCH{\scriptsize II}
regions. \citet{Pankonin2001} conducted a \chthreecn\,(12\,--\,11)
survey for hot cores and also found a correlation between the presence
of \chthreecn~and UCH{\scriptsize II} regions. In one of the few high
resolution studies \citet{Remijan2004} used the BIMA array to image
the hot core regions W51e1 and W51e2 and found that the
\chthreecn~emission was centred on known UCH{\scriptsize II} regions.

That we detect \chthreecn~towards isolated methanol masers, for which
there are no other tracers indicative of star-formation, clearly
suggests that these objects are internally heated. The lack of
\chthreecn~towards some isolated maser sources is consistent with them
being at a less advanced stage of evolution compared to
UCH{\scriptsize II} regions. In young hot cores we would expect any
emission to be concentrated in the central regions, where the
protostar has heated the dust sufficiently for icy mantles to
evaporate and for \chthreecn~to form. As the temperature rises icy
mantles continue to evaporate and the emitting region expands outwards
and becomes easier to detect. UCH{\scriptsize II} regions represent
the most advanced evolutionary stage before the young star emerges
from its natal cocoon. It is reasonable to assume a relatively
extended envelope of \chthreecn~may exist around the UCH{\scriptsize
  II} region, which is less beam diluted and easier for a single-dish
survey to detect. Such a scenario is lent weight by high resolutions 
($\sim$\,1\arcsec) observations of the hot core in
G29.96$-$0.02. \citet{Olmi2003} derive a kinetic temperature of 
$\sim$\,150\,K from vibrationally excited \chthreecn, significantly
higher than the $\sim$\,90\,K found by \citet{Pratap1999}, using
ground state lines. \citet{Olmi2003} also find evidence for a
temperature gradient, as the emission from higher energy transitions is
confined to increasingly compact regions.

Alternatively, the dominant \chthreecn~emission may emanate from an
unresolved hot core within the same beam. G29.96$-$0.02 is the classic
example of a hot core on the edge of a cometary UCH{\scriptsize II}
region (c.f. \citealt{Cesaroni1998}). We note that in the present
survey the \chthreecn~emission is classified as being associated with
the UCH{\scriptsize II} region, as at 3-mm wavelengths, Mopra does not
have the resolution to disentangle the hot core from the nearby
($\sim$\,5\,\arcsec) radio emission. Conclusions drawn from this work
refer to the properties of ``clumps'', which are may contain
more than one core. We speculate that clumps containing
UCH{\scriptsize II} regions are more likely to contain an evolved hot
core and thus have greater abundances of daughter species, like
\chthreecn.

Further enhancements come from the work of
\citet*{Mackay1999} who developed a chemical model of the hot core
G34.3$+$0.15. The model predicts an enhanced abundance of
\chthreecn~in the presence of a far-ultraviolet radiation
field. MacKay assumed a spherically symmetric core surrounded by a
photon dissociation region (PDR), created by the ultraviolet radiation
from a nearby OB association. It is more likely that the cores are
inhomogeneous and have clumpy density distributions, in which case the
incident radiation will penetrate deeper and the \chthreecn~abundance
will be further enhanced. Seven out of the nineteen UCH{\scriptsize
  II} regions in our sample are exactly coincident with methanol maser
sites, however the remaining twelve are offset by less than a beam and
it is likely that these sources have a complex clumpy structure.
 

\subsection{V$_{\rm LSR}$ and Linewidths}
\begin{figure}
  \begin{center}
    \includegraphics[height=6.9cm, angle=-90, trim=0 -0 -0 0]{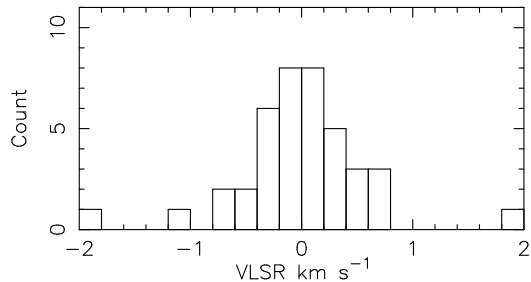}
    \caption{~V${\rm _{LSR}}$ (\chthreecn~--~\hthirteencop). The
      difference in velocity between the \chthreecn~and
      \hthirteencop~lines is within the errors, making it likely that
      the emission arises from the same source. Velocity errors are
      derived from Gaussian fits and are typically $\pm$\,0.4\,\kms.} 
    \label{fig:veldiff}
  \end{center}
\end{figure}

The difference between the \chthreecn~and \hthirteencop~V$_{\rm LSR}$
is plotted in Figure~\ref{fig:veldiff}. The velocity offsets are all
less than 2\,\kms, indicating that the molecules are probably within
the same star forming region. 92\,\% of the sources have velocity
offsets less than the fitting errors ($\pm$\,0.4\,\kms), suggesting
that they trace the same clump or core within the beam.

\begin{figure}
  \begin{center}
    \includegraphics[height=6.9cm, angle=-90, trim=0 -0 -0 0]{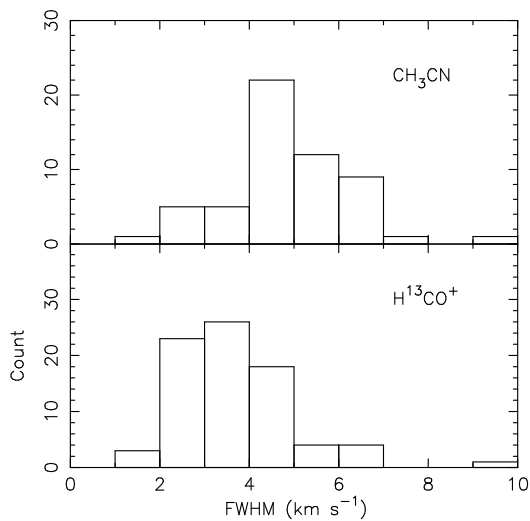}
    \caption{~\chthreecn\,(5\,--\,4) (top panel) and
    \hthirteencop\,(1\,--\,0) (bottom panel) linewidth
    distributions. Both distributions are roughly symmetrical about
    means of 4.9 and 3.5\,\kms~respectively.}
    \label{fig:vlsr_ch3cn_h13cop}
  \end{center}
\end{figure}

Figure~\ref{fig:vlsr_ch3cn_h13cop} displays the distributions of
linewidth from the Gaussian fits to the \chthreecn~and
\hthirteencop~lines. Both distributions are roughly symmetrical about
means of 4.9 and 3.5\,\kms respectively, higher than typical
linewidths of $<$\,2\,\kms~measured in low-mass star-forming
regions. At 100\,K the thermal linewidths of \chthreecn\,(5\,--\,4)
and \hthirteencop\,(1\,--\,0) are 0.20\,\kms~and
0.24\,\kms~respectively, with turbulence or bulk gas motions
accounting for most of the broadening. The \chthreecn~linewidth is
broader than the \hthirteencop~linewidth, which suggests that the two
species trace different spatial scales. This is consistent with
\chthreecn~being concentrated around a dynamic core and the
\hthirteencop~tracing a more extended quiescent envelope. 

We also note that sources with associated radio continuum emission
have larger linewidths than sources without (i.e. isolated maser
sources). The mean \chthreecn\,(5\,--\,4) linewidths are 6.2\,\kms~for
UCH{\scriptsize II} regions versus 4.7\,\kms~for isolated masers,
while the mean \hthirteencop\,(1\,--\,0) linewidths are 4.0\,\kms~for
UCH{\scriptsize II} regions versus 3.4\,\kms~for isolated
masers. Linewidths are a rough indicator of star-formation activity:
we expect to see greater linewidths towards more dynamic regions. The
greater linewidths reported for UCH{\scriptsize II} regions are
certainly consistent with a scenario where the isolated maser sources
are precursors to the UCH{\scriptsize II} phase.


\subsection{\hcop~Line Profiles}
\begin{figure}
  \begin{center}
    \includegraphics[width=7.9cm, trim=0 -0 -0 0]{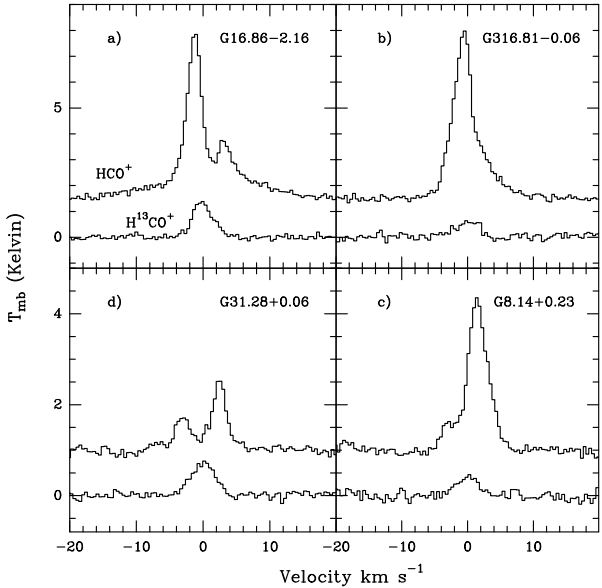}
    \caption{~Example \hcop~line profiles. {\bf a}) Blue double-peaked
    profile, {\bf b}) blue skewed profile, {\bf c}) red skewed
    profile, {\bf d}) red double-peaked profile.}
    \label{fig:profile_examples}
  \end{center}
\end{figure}
\begin{figure}
  \begin{center}
    \includegraphics[width=6.9cm, trim=0 -0 -0 0]{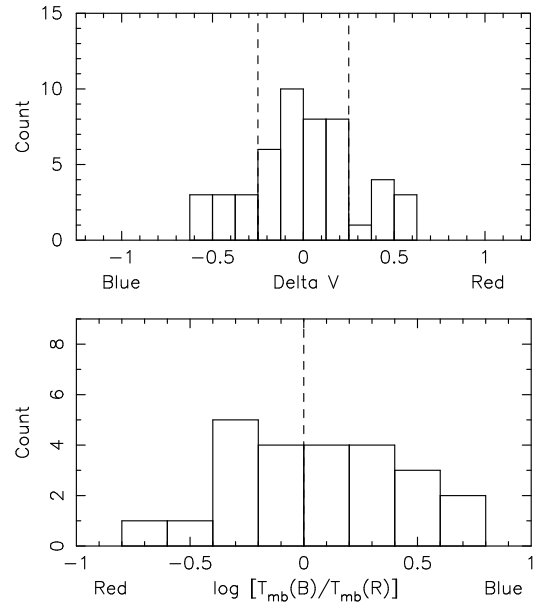}
    \caption{Distributions of red and blue profiles measured by
    ${\rm (v_{thick}-v_{thin})/v_{thin}}$ (upper panel) and
    ${\rm[T_{MB}(B)/T_{MB}(R)]}$ (lower panel). In the top panel the
    vertical dashed lines mark the absolute velocity difference beyond
    which profiles are considered to be blue or red skewed. In the
    lower panel the line divides red and blue profiles. One red and
    one blue source were discarded as the difference in peak height
    was within the 1\,$\sigma$ noise on the spectrum.}  
    \label{fig:profile_distrib}
  \end{center}
\end{figure}

Almost all \hcop~lines exhibit asymmetric line profiles or
high-velocity line wings. Complex profiles may be interpreted as
either multiple emitting regions along the same line of sight, or as a
single emitting region with cold absorbing gas intervening. By
examining the line profiles of optically thin \hthirteencop~we attempt
to distinguish between these two causes, and find that most sources
are composed of a single broad line with a self absorption
dip. Depending on the shape of the profile, we can infer inward or
outward motions. Blue-skewed profiles are predicted by collapse
models of star formation, however, rotation or outflow-blobs can also
produce similar line shapes. The presence of a statistical excess of
blue profiles in a survey can indicate that inflow is a likely
explanation \citep*{WuEvans2003}. Two methods of characterising line
profiles appear in the literature. Where the opacity is high and the
line takes on a double-peaked form, \citet{WuEvans2003} measure the
ratio of the blue to the red peak ${\rm [T_{MB}(B)/T_{MB}(R)]}$. A
blue profile fulfils the criterion: ${\rm [T_{MB}(B)/T_{MB}(R)]>1}$
by a statistically significant amount. At lower optical depths the
absorption will be less severe and the line will appear as a skewed
peak with a red or blue shoulder. \citet{Mardones1997}  suggest an
alternative measurement: the line may be classed as blue  if the
difference between the peak velocity of the optically-thick  and
optically-thin lines is greater than 1/4 the line width of the
optically thin line: ${\rm \delta
  v=(v_{thick}-v_{thin})/v_{thin}<-0.25}$. Similarly for a red
profile:  ${\rm \delta v=(v_{thick}-v_{thin})/v_{thin}>0.25}$.
Figure~\ref{fig:profile_examples} displays a range of \hcop~profiles
from blue to red. We have classed all detected \hcop~lines using one 
of the two schemes above. The results are displayed in
Figure~\ref{fig:profile_distrib}. We identify 9 blue and 8 red
profiles using the ${\rm \delta v}$ measurement, and 12 blue and 10
red profiles using the brightness method. To effect a comparison with
other surveys we adopt the concept of ``excess'' introduced by
\citet{Mardones1997} and later used by \citet{WuEvans2003}: ${\rm
  E=(N_{blue}-N_{red})/N_{total}}$, where ${\rm N_{total}}$ is the
number of sources in the sample. The excesses using the ${\rm \delta
  v}$ and brightness methods are E = 0.02 and 0.08, respectively. In
comparison \citet{WuEvans2003} measure excesses of 0.29 and 0.21 in a
sample of 28 low-mass star-forming regions. 

We also split our sample into radio-strong and radio-weak population,
searching for differences in the incidence of blue or red
profiles; however we do not find any convincing difference between the
populations.


\subsection{MSX colours}
\begin{figure*}
  \begin{center}
    \includegraphics[height=6.9cm, trim=0 -0 -0 0]{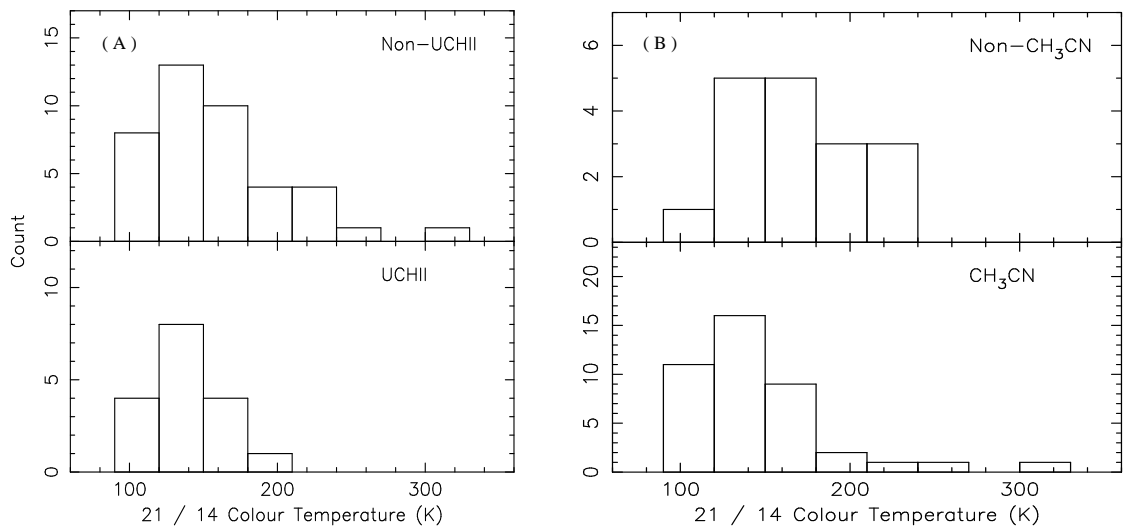}
    \caption{~Colour temperatures derived from the
    21\,/\,14\,\micron~ratio, for sources with and without associated
    UCH{\scriptsize II} regions, (A), and \chthreecn~emission, (B). A
    KS-test cannot distinguish between the distributions with
    confidence.}  
    \label{fig:hist_21_14}
  \end{center}
\end{figure*}

The MSX colours of our sample are primarily determined by the
temperature and optical depth of the dust in which they are
embedded. As a first attempt to distinguish between classes of source
we have derived the colour temperature from the 21/14\,\micron~ratio,
assuming blackbody emission. Larger ratios are equivalent to lower
temperatures and increasingly reddened colours. We have avoided using
the 21/8\,\micron~or 21/12\,\micron~ratios because of possible
contamination from PAHs emission at 7.7, 8.6 and 13.3\,\micron, and
silicate features at 9.7 and 11.3\,\micron. Distributions of derived
temperatures for sources split into classes with and without
UCH{\scriptsize II} regions and \chthreecn~emission are displayed in
Figure~\ref{fig:hist_21_14}. A KS-test fails to find any significant
difference between the classes.

\begin{figure}
  \begin{center}
    \includegraphics[height=7.9cm, angle=-90, trim=0 -0 -0 0]{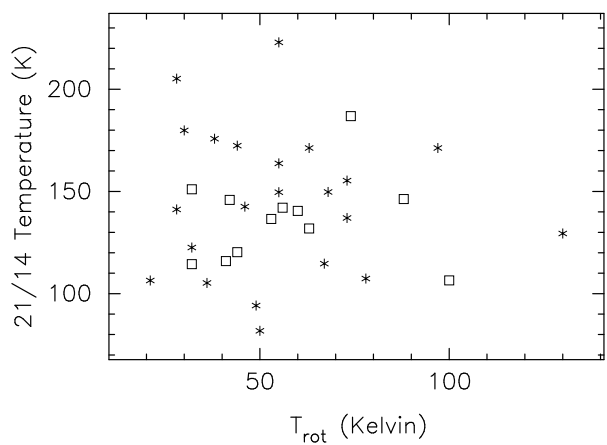}
    \caption{~21\,/\,14 \micron~ratio vs ${\rm T_{rot}}$. Squares
      represent sources containing UCH{\scriptsize II} regions. The
      21\,/\,14\,\micron~ratio does not appear to be correlated with the
      temperature obtained from the rotational analysis of the
      \chthreecn. Note also that the distribution of T$_{\rm rot}$ is
      identical for the UCH{\scriptsize II} regions and isolated
      masers.} 
    \label{fig:xy_21_14_trot_uchii}
  \end{center}
\end{figure}

Figure \ref{fig:xy_21_14_trot_uchii} shows a plot of colour
temperature against ${\rm T_{rot}}$, however no correlation is
evident, most likely because the thermal infrared and
\chthreecn~emission arise in different regions. Although optical depth
effects render colour ratios a poor indicator of kinetic temperature,
we note that the rotational temperatures are on average three times
lower than the colour temperatures.

\citet{Lumsden2002} have made a study of sources in the MSX catalogue
in an effort to perform a census of massive protostars in the Galactic
plane. They have arrived at a set of mid-infrared colour criteria
forming the first step in a selection process designed to find the
majority of massive protostars present in the Galaxy. As most massive
protostars have a featureless red continuum rising between 1 and
100\,\micron, they require that ${\rm F_8<F_{14}<F_{21}}$. Known
massive protostars, when plotted on a mid-infrared colour-colour
diagram, have an infrared excess such that ${\rm F_{21}/F_{8}>2}$. In
the absence of other selection criteria, sources which satisfy these
colour-cuts will include massive protostars, evolved stars, planetary
nebula and UCH{\scriptsize II} regions. Applying these criteria to our
data-set we initially discard  28 sources which have incomplete
data. Figure~\ref{fig:colour-colour} shows the mid-infrared
colour-colour diagrams for the remaining 55 sources. Associations with
UCH{\scriptsize II} regions and \chthreecn~are noted by way of
different symbols and the selection limits are marked as horizontal
and vertical dashed lines. Five sources fail the selection criteria:
G23.26$-$0.24 fails both colour-cuts, G30.90$+$0.16 fails because
${\rm F_{21}/F_{8}<2}$ and the remaining three (G6.54$-$0.11,
G12.72$-$0.22, G16.59$-$0.05) fail because ${\rm F_{14}<F_{8}}$. All
sources are radio-quiet and three have detected \chthreecn~emission,
but are unremarkable in other ways. No clustering is seen on the
colour-colour plots that would distinguish the classes of source,
except perhaps a slight enhancement of the 14/12\,\micron~ratio for
UCH{\scriptsize II} regions. UCH{\scriptsize II} regions are on
average also a factor of two brighter in all bands than radio-quiet
sources. To further separate young massive stars from planetary nebula
and evolved stars \citet{Lumsden2002} resorted to near-infrared JHK
colour selection criteria. 6.7\,GHz methanol-masers have not, to date,
been found towards low-mass star forming regions and we believe that
all of our maser sources represent massive stars \citep{Minier2005}.

Potentially the most interesting sources are `MSX-dark clouds', seen
in absorption against the Galactic plane, and the `MSX-red' sources,
detected only at 21\,\micron. It has been suggested that these sources
harbour massive protostars at earlier, more obscured phases of massive
star-formation, characterised by low temperatures (T$\sim$\,30\,K) and
spectral energy distributions which peak in the sub-millimetre and
far-infrared. Table~\ref{tab:msx_red_dark} summarises the properties
of the sources which fall into these categories. The `MSX-dark'
category contains thirteen sources, one of which (G30.71$-$0.06) is
within 30$''$ of a UCH{\scriptsize II} region. The `MSX-red' category
contains seven sources, one of which (G30.79$+$0.20) is also seen in
absorption at 8\,\micron~but is detected in emission at 21\,\micron.

We find that the `MSX-red' sources are indistinguishable from the bulk
of the sample, having approximately the same molecular detection rates
and brightness. Four out of eight sources exhibit asymmetric
\hcop~profiles, three of which are skewed to the red, suggestive of
outflowing motions.  
\begin{figure*}
  \begin{center}
    \includegraphics[height=6.9cm, trim=0 -0 -0 0]{figs/fig_16.epsi}
    \caption{~Mid-infrared colour-colour diagrams for sources detected
      in all bands. The following symbols mark classes of source:
      Square = UCH{\scriptsize II} region with \chthreecn, Triangle =
      UCH{\scriptsize II} without \chthreecn, Star =
      Non-UCH{\scriptsize II} region with \chthreecn, Cross =
      Non-UCH{\scriptsize II} region without \chthreecn. The dashed
      lines mark the selection criteria proposed by
      \citet{Lumsden2002}: ${\rm F_8<F_{14}<F_{21}}$ and ${\rm
      F_{21}/F_{8}<2}$. Sources which fulfil these criteria include
      MYSOs, UCH{\scriptsize II} regions, evolved stars and planetary
      nebula.}  
    \label{fig:colour-colour}
  \end{center}
\end{figure*}

`MSX-dark' sources exhibit distinct differences from the bulk of the
sample. 91\,\% were detected in \chthreecn, and the \chthreecn~and
\hthirteencop~brightness temperatures tend to be higher. However, by
their nature dark-clouds are located nearby making it easier to detect
weak lines. Seven of twelve sources exhibit asymmetric \hcop~profiles,
six of which are skewed to the blue, indicative of in-falling motions,
and consistent with a scenario of cloud collapse. Further analysis is
required to judge if these cores will collapse to form massive stars. 


\subsection{Maserless Sources}
Six maserless sources, identified by thermal emission at 1.2-mm from
the work of \citet{Hill2005}, were included in the source list as
potential precursors to the hot core phase. One of these,
G5.89$-$0.39, was subsequently identified as an UCH{\scriptsize II}
region. The remaining five are spread among the classes of source
identified in previous section: G0.26$+$0.01 and G14.99$-$0.70 are
MSX-dark clouds, G5.90$-$0.44 and G12.72$-$0.22 are bright and
partially resolved MSX sources, and G15.03$-$0.71 is embedded in
confused region of extended emission near the M17 molecular cloud. All
lie at distances nearer than 4\,kpc, except for G0.26+0.01 which we
place at the distance of the Galactic centre. \chthreecn~is moderately
detected towards all sources, except G15.03-0.71. \hcop~and
\hthirteencop~are found towards all sources while the
\hthirteencop~linewidths vary from 2\,--\,4.5\,\kms~without any
obvious relation to MSX brightness or class. The absence of a further
link between these maserless sources and the spread in observed
properties suggests a range of evolutionary phases; i.e. the absence
of maser emission towards these sources does not necessarily indicate
an early phase star-formation, even with the presence of \chthreecn.


\section {Summary \& Conclusions}
We have detected 58 candidate hot molecular cores through the presence
of \chthreecn~emission from a sample of 83 methanol maser selected
star-forming regions, including 43 new detections. All sites are
associated with mm-continuum thermal emission, seen at 1.2-mm,
450\,\micron~and 850\,\micron. Six sites are not directly associated
with maser emission but are within the same star-forming region. Our
major findings are as follows: 
\begin{enumerate}
\item \chthreecn~is commonly detected towards methanol maser sites,
  but is more prevalent and brighter in the presence of a
  UCH{\scriptsize II} region, independant of the distance to the
  source.
\item \chthreecn~is detected towards isolated methanol maser sites,
  where no nearby external heating sources exist. This strongly
  suggests that these sources are internally heated. Conversely, the
  lack of \chthreecn~towards some isolated maser sites is consistent
  with them being at a less advanced stage of evolution than
  UCH{\scriptsize II} regions.
\item We report the \chthreecn~rotational temperatures for 38 sources
  where at least three K components were detected. Values range from
  28 to 166\,K. These values are generally lower than those found in
  previous studies utilising higher energy transitions. Derived column
  densities are poorly constrained but are comparable to previous
  studies. There is no significant difference between the isolated
  maser and UCH{\scriptsize II} populations.
\item \hcop~and \hthirteencop~are detected towards 99\,\% and 98\,\%
  of the sample, respectively. Low excitation temperatures derived
  from \hcop~lead us to believe that the emission is generally beam
  diluted in our 36\,\arcsec~beam, however, self absorption in the
  \hcop~line profile may also cause us to underestimate the line
  brightness. We derived beam averaged column densities assuming a
  [\hcop]\,/\,[\hthirteencop] ratio of 50 and an excitation
  temperature of 15\,K.  
\item Most \hcop~line profiles exhibit asymmetries due to
  self-absorption, which may be interpreted as inward or outward
  motions. Approximately equal numbers of red and blue profiles are
  found towards UCH{\scriptsize II} regions, isolated masers and
  maserless sources.
\item The majority of sources with multi-colour MSX fluxes pass the
  MYSO colour selection criteria of \citet{Lumsden2002}. Five maser
  sources fail the colour cuts and are likely very young massive
  star-forming cores. Sources with very red MSX colours are
  indistinguishable from the bulk of the sample in terms of
  \chthreecn~detections and brightness. MSX-dark clouds are notable as
  \chthreecn~is generally brighter than in sources detected in
  emission with MSX; however this may be as a result of their near
  distances. Where asymmetric \hcop~profiles are seen towards MSX-dark
  clouds they tend to be blue-skewed, indicative of inward motions.
\end{enumerate}

In continuation of this work we will report on the detection of other
molecules in a forthcoming paper. We have surveyed the same sources
for CH$_3$OH\,(2\,--\,1), $^{13}$CO\,(1\,--\,0),
N$_2$H$^+$\,(1\,--\,0), HCN\,(1\,--\,0) and HNC\,(1\,--\,0), also
using the Mopra antenna at 3-mm.


\section*{Acknowledgements}
The authors wish to thank the staff of the ATNF for their tireless
support of the Mopra telescope since 1999. The Mopra telescope is
operated through a collaborative arrangement between the University of
New South Wales and the CSIRO. We also wish to thank the Australian
research council and UNSW for grant support. CRP was supported by a
School of Physics Scholarship during the course of his PhD. EFL
gratefully acknowledges support from the ATNF Distinguished Visitor
Program, and US National Science Foundation grant AST03-0750.
We are also grateful to the anonymous referee for comments and
discussions that helped improve the science and presentation.



\newpage
\clearpage
\begin{table}
  \begin{center}
    \begin{minipage}{95mm}
      \caption{~Details of observed transitions}\label{tab:transitions}
      \begin{tabular}{lcccccc}
	\hline
	Species       & Transition\,$^{\alpha}$ & Frequency & ${\rm E_{u}/k}$ & ${\rm A_{ul}}$\,$^{\beta}$ & ${\rm g_{JK}}$\,$^{\gamma}$\\
                      &                         & (GHz)     &    (K)          &(${\rm \times 10^{-5}\,s^{-1}}$) \\
	\hline  
	\chthreecn    & $5_{0}\rarr 4_{0}$ & ~~91.987054 & ~~13.24  & ~~6.121  & ~~44\\
                      & $5_{1}\rarr 4_{1}$ & ~~91.985284 & ~~20.39  & ~~5.875  & ~~44\\
      	              & $5_{2}\rarr 4_{2}$ & ~~91.980000 & ~~41.82  & ~~5.139  & ~~44\\
	              & $5_{3}\rarr 4_{3}$ & ~~91.971374 & ~~77.53  & ~~3.913  & ~~88\\
	              & $5_{4}\rarr 4_{4}$ & ~~91.959206 & 127.51   & ~~2.200  & ~~44\\
                      & $6_{0}\rarr 5_{0}$ & 110.383494  &  ~~18.54 &  10.895  & ~~52\\
                      & $6_{1}\rarr 5_{1}$ & 110.381376  &  ~~25.68 &  10.592  & ~~52\\
      	              & $6_{2}\rarr 5_{2}$ & 110.374968  &  ~~47.11 & ~~9.681  & ~~52\\
	              & $6_{3}\rarr 5_{3}$ & 110.364470  &  ~~82.82 & ~~8.166  & 104\\
	              & $6_{4}\rarr 5_{4}$ & 110.349760  &   134.25 & ~~6.045  & ~~52\\
	\hcop	      & $1\rarr0$          & ~~89.188526 & ~~~4.28  & ~~3.02   & ~~~3\\
	\hthirteencop & $1\rarr0$          & ~~86.754330 & ~~~4.16  & ~~2.80   & ~~~3\\
	\hline
      \end{tabular}
      \begin{flushleft}
	$^{\alpha}$~${\rm J_K}$ quantum numbers.\\	
	\smallskip
	$^{\beta}$~Einstein A coefficients for symmetric tops and
	linear rotors are given by ${\rm
	(16\pi^3\nu^3\mu^2S)/(3\epsilon_{0}hc^3g_u)}$ where ${\rm
	g_u=(2J+1)}$ and $\mu$ is the electric dipole
	moment for the molecule. Other constants take their usual
	values and are in SI units.\\
	\smallskip
	$^{\gamma}$~${\rm g_{jk}=g_u\,S(I,J)}$ is the degeneracy of
	the ${\rm J_K}$ rotational level as given by
	\citet{Araya2005}, matched with the partition function quoted
	in the same paper. ${\rm S(I,J) = (J^2-K^2)/J}$ is the
	intrinsic line strength.\\
      \end{flushleft}
    \end{minipage}
  \end{center}
\end{table}

\newpage
\clearpage
\begin{table*}
  \begin{center}
    \begin{minipage}{160mm}
      \caption{~Details of the 83 sources observed.}\label{tab:sources}
      \begin{tabular}{lcccccccccc}
	\hline
	Galactic & Right     & Declination &  Velocity &  Adopted\,$^{\alpha}$ & Luminosity\,$^{\beta}$ & Single\,$^{\gamma}$ & \multicolumn{3}{c}{Other\,$^{\delta}$} & Molecular\\
	Name     & Ascension &             &  (LSR)    &  Distance  & & Star   & \multicolumn{3}{c}{Associations}  & Cloud\\
        & (J2000)   & (J2000)     &  (\kms)   &  (kpc)     & ($\times$10$^4$L$_{\odot}$) & Type   &&&& Name           \\
	\hline 
	G0.21+0.00 &  17:46:07.7 & $-$28:45:20 & 44.6 & 8.4$^t$ & 11.9 & O8 & M & R & MIR \\
	G0.26+0.01 &  17:46:11.3 & $-$28:42:48 & 26.0 & 8.4$^t$ & -- & -- & M & -- & DRK \\
	G0.32$-$0.20 &  17:47:09.1 & $-$28:46:16 & 18.9 & 8.5$^t$ & 39.9 & O6 & M & R & MIR \\
	G0.50+0.19 &  17:46:04.0 & $-$28:24:51 & $-$6.1 & 2.6$^p$ & 0.9 & B1 & M & -- & MIR \\
	G0.55$-$0.85 &  17:50:14.5 & $-$28:54:31 & 17.5 & 8.5$^t$ & 85.5 & O5 & M & R & MIR & RCW142\\
	G0.84+0.18 &  17:46:52.8 & $-$28:07:35 & 5.9 & 5.6 & 3.7 & O9.5 & M & -- & MIR \\
	G1.15$-$0.12 &  17:48:48.5 & $-$28:01:12 & $-$17.2 & 8.5 & 10.9 & 08 & M & -- & MIR \\
	G2.54+0.20 &  17:50:46.5 & $-$26:39:45 & 10.1 & 4.4$^r$ & -- & -- & M & -- & DRK \\
	G5.89$-$0.39 &  18:00:31.0 & $-$24:03:59 & 9.3 & 2.0$^a$ & 14.7 & O7.5 & M & R & MIR & W28\\
	G5.90$-$0.43 &  18:00:40.9 & $-$24:04:21 & 7.0 & 2.0$^d$ & 6.8 & O9.5 & M & -- & MIR & W28\\
	G5.90$-$0.44 &  18:00:43.9 & $-$24:04:47 & 9.5 & 2.5$^d$ & 3.7 & B0 & M & -- & MIR & W28\\
	G6.54$-$0.11 &  18:00:50.9 & $-$23:21:29 & 0.0 & 14.9$^w$ & 37.2 & 06 & M & -- & MIR \\
	G6.61$-$0.08 &  18:00:54.0 & $-$23:17:02 & 7.7 & 14.9 & 5.7 & O9.5 & M & -- & MIR \\
	G8.14+0.23 &  18:03:00.8 & $-$21:48:10 & 19.2 & 3.3$^w$ & 8.5 & O8.5 & M & R & MIR & W30\\
	G8.67$-$0.36 &  18:06:19.0 & $-$21:37:32 & 34.8 & 4.5$^d$ & 7.4 & O9 & M & R & MIR \\
	G8.68$-$0.37 &  18:06:23.5 & $-$21:37:11 & 37.2 & 4.8$^d$ & -- & -- & M & -- & DRK \\
	G9.62+0.19 &  18:06:14.8 & $-$20:31:37 & 4.4 & 5.7$^{ho}$ & 28.0 & O6.5 & M & R & MIR \\
	G9.99$-$0.03 &  18:07:50.1 & $-$20:18:57 & 48.9 & 5.0 & 2.0 & BO.5 & M & -- & MIR \\
	G10.10$-$0.72 & 18:05:18.2 & $-$19:51:14 & -4.0 & 16 & -- & -- & M & -- & -- \\
	G10.29$-$0.13 &  18:08:49.4 & $-$20:05:59 & 13.7 & 2.2$^d$ & 1.1 & B1 & M & -- & MIR \\
	G10.30$-$0.15 &  18:08:55.5 & $-$20:05:58 & 13.0 & 2.1$^d$ & 6.6 & O9.5 & M & R & MIR & W31\\
	G10.32$-$0.16 &  18:09:01.5 & $-$20:05:08 & 12.2 & 2.2$^d$ & 8.1 & 09 & M & -- & MIR & W31\\
	G10.34$-$0.14 &  18:09:00.0 & $-$20:03:36 & 12.2 & 2.2$^d$ & 1.7 & B0.5 & M & -- & MIR \\
	G10.44$-$0.02 &  18:08:44.9 & $-$19:54:38 & 75.4 & 6.0$^r$ & -- & -- & M & -- & DRK \\
	G10.47+0.03 &  18:08:38.2 & $-$19:51:50 & 67.0 & 5.7$^{ce}$ & 23.0 & O6.5 & M & R & MIR \\
	G10.48+0.03 &  18:08:37.9 & $-$19:51:15 & 66.2 & 5.7$^{ce}$ & -- &--  & M & -- & DRK \\
	G10.63$-$0.33 &  18:10:18.0 & $-$19:54:05 & $-$4.1 & 6.0$^d$ & 56.8 & 05.5 & M & -- & MIR & W31\\
	G10.63$-$0.38 &  18:10:29.2 & $-$19:55:41 & $-$3.1 & 6.0$^d$ & 26.4 & O6.5 & M & R & MIR \\
	G11.50$-$1.49 &  18:16:22.1 & $-$19:41:28 & 10.3 & 1.6 &  0.5 & B2 & M & -- & MIR \\
	G11.94$-$0.15 &  18:12:17.3 & $-$18:40:03 & 42.6 & 4.4 & 0.6 & B2 & M & -- & MIR \\
	G11.94$-$0.62 &  18:14:00.9 & $-$18:53:27 & 37.9 & 4.1$^{ch}$ & 7.6 & O9 & M & R & MIR \\
	G11.99$-$0.27 &  18:12:51.2 & $-$18:40:40 & 59.7 & 5.2 & 1.2 & B1 & M & -- & MIR \\
	G12.03$-$0.03 &  18:12:01.9 & $-$18:31:56 & 110.5 & 6.7 & 1.3 & B0.5 & M & -- & MIR \\
	G12.18$-$0.12 &  18:12:41.0 & $-$18:26:22 & 26.5 & 13.4$^d$ & -- & -- & M & -- & MIR \\
	G12.21$-$0.09 &  18:12:37.5 & $-$18:24:08 & 23.8 & 13.6$^d$ & -- & -- & M & -- & MIR \\
	G12.68$-$0.18 &  18:13:54.7 & $-$18:01:41 & 56.5 & 4.9$^s$ & 1.5 & B0.5 & M & -- & MIR & W33 \\
	G12.72$-$0.22 &  18:14:07.0 & $-$18:00:37 & 34.2 & 3.7$^s$ & 4.7 & B0 & M & -- & MIR & W33\\
	G12.89+0.49 &  18:11:51.4 & $-$17:31:30 & 33.3 & 3.6$^m$ & 3.4 & B0 & M & -- & MIR \\
	G12.91$-$0.26 &  18:14:39.5 & $-$17:52:00 & 37.0 & 3.9$^s$ & 7.3 & O9.5 & M & -- & MIR & W33\\
	G14.60+0.02 &  18:17:01.1 & $-$16:14:39 & 24.7 & 2.8$^d$ & 1.0 & B1 & M & R & MIR \\
	G14.99$-$0.70 &  18:20:23.1 & $-$16:14:43 & 18.7 & 1.6$^h$ & -- & -- & M & -- & DRK & M17\\
	G15.03$-$0.68 &  18:20:24.8 & $-$16:11:35 & 19.5 & 1.6$^h$ & 16.3 & O7 & M & -- & MIR & M17\\
	G15.03$-$0.71 &  18:20:30.3 & $-$16:12:43 & 21.0 & 1.6$^h$ & 3.1 & B0 & M & -- & MIR & M17\\
	G16.59$-$0.05 &  18:21:09.1 & $-$14:31:49 & 59.9 & 4.6$^{co}$ & 2.5 & B0.5 & M & -- & MIR \\
	G16.86$-$2.16 &  18:29:24.4 & $-$15:16:04 & 17.8 & 1.9 & 0.4 & B2 & M & -- & MIR \\
	G19.36$-$0.03 &  18:26:25.2 & $-$12:03:53 & 26.7 & 2.5 & 0.5 & B2 & M & -- & MIR \\
	G19.47+0.17 &  18:25:54.7 & $-$11:52:34 & 19.7 & 1.9 & 1.3 & B0.5 & M & -- & MIR \\
	G19.49+0.15 &  18:26:00.4 & $-$11:52:22 & 23.0 & 2.2 & 0.6 & B2 & M & -- & MIR \\
	G19.61$-$0.13 &  18:27:16.4 & $-$11:53:38 & 56.9 & 4.2$^d$ & 2.1 & B0.5 & M & -- & MIR \\
	G19.70$-$0.27 &  18:27:55.9 & $-$11:52:39 & 43.0 & 3.5 & 0.6 & B2 & M & -- & MIR \\
	G21.88+0.01 &  18:31:01.7 & $-$09:49:01 & 23.3 & 2.0 & -- & -- & M & -- & -- \\
	G22.36+0.07 &  18:31:44.1 & $-$09:22:13 & 84.1 & 10.5$^s$ & 7.9 & O9 & M & -- & MIR \\
	G23.26$-$0.24 &  18:34:31.8 & $-$08:42:47 & 61.5 & 4.2 & 2.6 & B0.5 & M & -- & MIR \\
	G23.44$-$0.18 &  18:34:39.2 & $-$08:31:32 & 101.6 & 5.9$^d$ & -- & -- & M & -- & DRK \\
	G23.71$-$0.20 &  18:35:12.4 & $-$08:17:40 & 69.0 & 4.5 & 1.6 & B0.5 & M & -- & MIR \\
	G24.79+0.08 &  18:36:12.3 & $-$07:12:11 & 110.5 & 7.7$^{be}$ & 20.4 & O6.5 & M & -- & MIR \\
	G24.85+0.09 &  18:36:18.4 & $-$07:08:52 & 108.9 & 6.3$^{co}$ & 10.8 & O8 & M & -- & MIR \\
	G25.65+1.05 &  18:34:20.9 & $-$05:59:40 & 42.4 & 3.1$^m$ & 2.5 & B0.5 & M & R & MIR \\
	\hline
      \end{tabular}
    \end{minipage}
  \end{center}
\end{table*}
\newpage
\clearpage
\begin{table*}
  \begin{center}
    \begin{minipage}{160mm}
      \contcaption{}
      \begin{tabular}{lcccccccccc}
	\hline
	Galactic & Right     & Declination &  Velocity &  Adopted\,$^{\alpha}$ & Luminosity\,$^{\beta}$ & Single\,$^{\gamma}$ & \multicolumn{3}{c}{Other\,$^{\delta}$} & Molecular\\
	Name     & Ascension &             &  (LSR)    &  Distance  &                             & Star   & \multicolumn{3}{c}{Associations} &Cloud \\
        & (J2000)   & (J2000)     &  (\kms)   &  (kpc)     & ($\times$10$^4$L$_{\odot}$) & Type   &&&& Name           \\
	\hline 
	G25.71+0.04 &  18:38:03.1 & $-$06:24:15 & 98.7 & 9.5$^d$ & 8.9 & O8.5 & M & -- & MIR \\
	G25.83$-$0.18 &  18:39:03.6 & $-$06:24:10 & 93.4 & 5.6$^r$ & -- & -- & M & -- & DRK \\
	G28.15+0.00 &  18:42:42.2 & $-$04:15:32 & 98.6 & 5.9 & 2.0 & B0.5 & M & -- & MIR \\
	G28.20$-$0.05 &  18:42:58.1 & $-$04:13:56 & 95.6 & 6.8$^r$ & 17.1 & O7 M & R & MIR \\
	G28.28$-$0.36 &  18:44:13.3 & $-$04:18:03 & 48.9 & 3.3$^{so}$ & -- & -- & M & R & -- \\
	G28.31$-$0.39 &  18:44:22.0 & $-$04:17:38 & 86.2 & 5.2$^{so}$ & 4.0 & B0 & M & -- & MIR \\
	G28.83$-$0.25 &  18:44:51.1 & $-$03:45:48 & 87.1 & 5.3$^{so}$ & 3.5 & B0 & M & -- & MIR \\
	G29.87$-$0.04 &  18:46:00.0 & $-$02:44:58 & 100.9 & 6.3 & 5.3 & B0 & M & -- & MIR \\
	G29.96$-$0.02 &  18:46:04.8 & $-$02:39:20 & 97.6 & 6.0$^p$ & 74.0 & O5 & M & R & MIR & W43S\\
	G29.98$-$0.04 &  18:46:12.1 & $-$02:38:58 & 101.6 & 6.3$^r$ & -- & -- & M & -- & DRK \\
	G30.59$-$0.04 &  18:47:18.6 & $-$02:06:07 & 41.8 & 11.8$^{so}$ & 32.6 & O6 & M & -- & MIR \\
	G30.71$-$0.06 &  18:47:36.5 & $-$02:00:31 & 92.8 & 5.8$^r$ & -- & -- & M & R & DRK & W43 \\
	G30.76$-$0.05 &  18:47:39.7 & $-$01:57:22 & 93.0 & 5.8$^r$ & 12.1 & O8 & M & -- & MIR & W43\\
	G30.78+0.23 &  18:46:41.5 & $-$01:48:32 & 41.8 & 2.9 & -- & -- & M & -- & MIR \\
	G30.79+0.20 &  18:46:48.1 & $-$01:48:46 & 81.6 & 5.1$^r$ & 2.3 & B0.5 & M & -- & DRK \\
	G30.82$-$0.05 &  18:47:46.5 & $-$01:54:17 & 96.6 & 6.0$^r$ & 2.7 & B0.5 & M & -- & DRK & W43\\
	G30.82+0.28 &  18:46:36.1 & $-$01:45:18 & 97.8 & 8.5$^l$ & -- & -- & M & -- & MIR \\
	G30.90+0.16 &  18:47:09.2 & $-$01:44:09 & 106.2 & 7.3$^t$ & 4.8 & B0 & M & -- & MIR \\
	G31.28+0.06 &  18:48:12.4 & $-$01:26:23 & 109.4 & 7.2$^t$ & 12.0 & O8 & M & R & MIR \\
	G31.41+0.31 &  18:47:34.3 & $-$01:12:47 & 96.7 & 7.9$^{ce}$ & 17.0 & O7 & M & R & MIR \\
	G316.81$-$0.06 &  14:45:26.9 & $-$59:49:16 & $-$38.7 & 2.7 & 13.2 & O7.5 & M & -- & MIR \\
	G318.95$-$0.20 &  15:00:55.3 & $-$58:58:54 & $-$34.5 & 2.4 & 2.9 & B0.5 & M & -- & MIR \\
	G323.74$-$0.26 &  15:31:45.8 & $-$56:30:50 & $-$49.6 & 3.3 & 5.3 & B0 & M & -- & MIR \\
	G331.28$-$0.19 &  16:11:26.9 & $-$51:41:57 & $-$88.1 & 5.4 & 23.0 & O6.5 & M & -- & MIR \\
	G332.73$-$0.62 &  16:20:02.7 & $-$51:00:32 & $-$50.2 & 3.5 & -- & -- & M & -- & DRK \\
	\hline
      \end{tabular}
      \begin{flushleft}
$^{\alpha}$~The distance given in column 5 is the assumed kinematic
distance, in kpc, based on the rotation curve of
\citet{BrandBlitz1993}. We have resolved the kinematic distance
ambiguity for 58 sources, mainly using the work of the following
authors: 
$^{a}$\,\citet{Acord1998}, $^{ho}$\,\citet{Hofner1994},
$^{be}$\,\citet{Beltran2005}, $^{p}$\,\citet{Pratap1999},
$^d$\,\citet{Downes1980}, $^w$\,\citet{Wink1982},
$^{ch}$\,\citet{Churchwell1990}, $^{ce}$\,\citet{Cesaroni1998},
$^{m}$\,\citet{Molinari1996}, $^{so}$\,\citet{Solomon1987},
$^{co}$\,\citet{Codella1997}, $^l$\,\citet{Larionov1999},
$^h$\,\citet{Hanson1997}, $^s$\,\citet{Stier1984}. 
Also, the near and far distances for sources marked with a `$t$' were
within 1\,kpc and they were placed at the middle distance. Sources
marked with a `$r$' were seen in absorption at 8\micron~(dark clouds)
and were placed at the near distance. The remaining 25 sources were
assumed to lie at the near distance except for G10.10$-$0.72.\\ 
\smallskip
$^{\beta}$~Bolometric luminosities were calculated from a two
component greybody fit to the SED containing SCUBA and MSX fluxes
\citep{Walsh2003}. The IRAS fluxes were included in the fit only if
known with confidence. Sources without a quoted luminosity had
insufficient data to constrain a fit.\\ 
\smallskip
$^{\gamma}$~Approximate spectral types for a given luminosity were
read from Table~1 of \citet{Panagia1973}.\\  
\smallskip
$^{\delta}$~`M' indicates a 6.7\,GHz methanol-maser at the
position. `R' indicates a UCH{\scriptsize II} region lies within the
Mopra beam. `MIR' indicates a detection of mid-infrared MSX emission
near the source, `DRK' indicates the source is associated with a MSX
dark cloud.
      \end{flushleft}
    \end{minipage}
  \end{center}
\end{table*}

\newpage
\clearpage
\begin{table*}
  \begin{center}
    \begin{minipage}{220mm}
    \caption{~Summary of detections towards observed sources.}\label{tab:detections}
      \begin{tabular}{lcccccclccccc}
	\hline
	Source & CH$_3$CN\,$^{\alpha}$  & HCO$^+$ &  H$^{13}$CO$^+$ &  21\micron\,$^{\beta}$ &  8\micron\,$^{\beta}$ & & Source & CH$_3$CN\,$^{\alpha}$ & HCO$^+$ &  H$^{13}$CO$^+$ &  21\micron\,$^{\beta}$ &  8\micron\,$^{\beta}$\\
	\cline{1-6} \cline{8-13}
G0.21+0.00 & n~~~-- & y & n & y & y & & G15.03-0.71 & n~~~-- & y & y & y & y\\
G0.26+0.01 & y~~~-- & y & y & d & d & & G16.59-0.05 & y~~~-- & y & y & y & y\\
G0.32-0.20 & y~~~-- & y & y & y & y & & G16.86-2.16 & y~~~y & y & y & y & y\\
G0.50+0.19 & n~~~-- & y & y & y & y & & G19.36-0.03 & y~~~-- & y & y & y & y\\
G0.55-0.85 & y~~~y & y & y & y & y & & G19.47+0.17 & y~~~n & y & y & y & y\\
G0.84+0.18 & y~~~n & y & y & y & y & & G19.49+0.15 & n~~~-- & y & y & y & y\\
G1.15-0.12 & y~~~n & y & y & y & y & & G19.61-0.13 & y~~~-- & y & y & y & y\\
G2.54+0.20 & n~~~-- & y & y & n & d & & G19.70-0.27 & n~~~-- & y & y & y & y\\
G5.89-0.39 & y~~~-- & y & y & y & y & & G21.88+0.01 & n~~~-- & y & y & n & n\\
G5.90-0.43 & y~~~y & y & y & y & y & & G22.36+0.07 & y~~~-- & y & y & y & y\\
G5.90-0.44 & y~~~-- & y & y & y & y & & G23.26-0.24 & n~~~-- & y & y & n & y\\
G6.54-0.11 & n~~~-- & y & n & y & y & & G23.44-0.18 & y~~~y & y & y & n & d\\
G6.61-0.08 & n~~~-- & y & y & y & y & & G23.71-0.20 & n~~~-- & y & y & y & y\\
G8.14+0.23 & y~~~-- & y & y & y & y & & G24.79+0.08 & y~~~y & y & y & y & n\\
G8.67-0.36 & y~~~-- & y & y & y & y & & G24.85+0.09 & n~~~-- & y & y & y & y\\
G8.68-0.37 & y~~~-- & y & y & n & d & & G25.65+1.05 & y~~~-- & y & y & y & y\\
G9.62+0.19 & y~~~n & y & y & y & y & & G25.71+0.04 & y~~~-- & y & y & y & n\\
G9.99-0.03 & y~~~-- & y & y & y & y & & G25.83-0.18 & y~~~y & y & y & n & d\\
G10.10+0.72 & n~~~-- & n & n & n & n & & G28.15+0.00 & y~~~-- & y & y & y & y\\
G10.29-0.13 & y~~~n & y & y & y & y & & G28.20-0.05 & y~~~-- & y & y & y & y\\
G10.30-0.15 & y~~~y & y & y & y & y & & G28.28-0.36 & y~~~-- & y & y & n & n\\
G10.32-0.16 & y~~~y & y & y & y & y & & G28.31-0.39 & n~~~-- & y & y & y & y\\
G10.34-0.14 & y~~~y & y & y & n & y & & G28.83-0.25 & y~~~-- & y & y & y & n\\
G10.44-0.02 & y~~~y & y & y & n & d & & G29.87-0.04 & n~~~-- & y & y & y & y\\
G10.47+0.03 & y~~~-- & y & y & y & y & & G29.96-0.02 & y~~~y & y & y & y & y\\
G10.48+0.03 & y~~~-- & y & y & n & d & & G29.98-0.04 & y~~~y & y & y & n & d\\
G10.63-0.33 & n~~~-- & y & y & y & y & & G30.59-0.04 & n~~~-- & y & y & y & y\\
G10.63-0.38 & y~~~y & y & y & y & y & & G30.71-0.06 & y~~~y & y & y & d & d\\
G11.50-1.49 & n~~~-- & y & y & y & y & & G30.76-0.05 & y~~~y & y & y & y & y\\
G11.94-0.15 & n~~~-- & y & y & y & y & & G30.78+0.23 & n~~~-- & y & y & n & y\\
G11.94-0.62 & y~~~y & y & y & y & y & & G30.79+0.20 & y~~~-- & y & y & y & d\\
G11.99-0.27 & n~~~-- & y & y & y & y & & G30.82-0.05 & y~~~y & y & y & n & d\\
G12.03-0.03 & n~~~-- & y & y & y & y & & G30.82+0.28 & n~~~-- & y & y & y & y\\
G12.18-0.12 & n~~~-- & y & y & y & n & & G30.90+0.16 & y~~~-- & y & y & y & y\\
G12.21-0.09 & n~~~-- & y & y & y & n & & G31.28+0.06 & y~~~y & y & y & y & y\\
G12.68-0.18 & y~~~y & y & y & y & y & & G31.41+0.31 & y~~~y & y & y & y & y\\
G12.72-0.22 & y~~~-- & y & y & y & y & & G316.81-0.06 & y~~~y & y & y & y & y\\
G12.89+0.49 & y~~~n & y & y & y & y & & G318.95-0.20 & y~~~n & y & y & y & y\\
G12.91-0.26 & y~~~y & y & y & y & y & & G323.74-0.26 & y~~~y & y & y & y & n\\
G14.60+0.02 & y~~~-- & y & y & y & n & & G331.28-0.19 & y~~~n & y & y & y & y\\
G14.99-0.70 & y~~~-- & y & y & n & d & & G332.73-0.62 & y~~~n & y & y & d & d\\
\cline{8-13}
G15.03-0.68 & y~~~y & y & y & y & y & & \% Detections$^{\gamma}$: & 70 & 99 & 98 & 78 & 72\\
	\hline
      \end{tabular}
    \end{minipage}
  \end{center}
  \begin{flushleft}
    $^{\alpha}$~33 sources were searched for both \chthreecn\,(5\,--\,4)
    (first entry), and \chthreecn\,(6\,--\,5) (second entry). \\
    \smallskip
    $^{\beta}$~MSX dark clouds seen in absorption against background
    radiation at 8 or 21\,\micron~are marked with a `d'.\\
    \smallskip
    $^{\gamma}$~If dark clouds are counted then the detection rate
    rises to 82\,\% and 88\,\% for the 21 and 8\,\micron~MSX bands
    respectively.\\
  \end{flushleft}
\end{table*}

\newpage
\clearpage
\begin{table*}
    \begin{center}
      \begin{minipage}{155mm}
      \caption{~\chthreecn~Line Parameters.}\label{tab:mclineparams}
      \begin{tabular}{lccccccc}
	\hline
	Source\,$^{\alpha}$  & V$_{\rm LSR}$\,$^{\beta}$ & $\delta$V & T$_{\rm MB}$~dv,~K\,=\,0 & T$_{\rm MB}$~dv,~K\,=\,1 & T$_{\rm MB}$~dv,~K\,=\,2 & T$_{\rm MB}$~dv,~K\,=\,3 & T$_{\rm MB}$~dv,~K\,=\,4 \\
	& (\kms) & (\kms)      & (K \kms)     & (K \kms)     & (K \kms)     & (K \kms)     & (K \kms)     \\
	\hline
G0.32$-$0.20 & 18.09 & 2.87 & 0.29 $\pm$ 0.13 & 0.25 $\pm$ 0.12 & -- & -- & --  \\
G0.55$-$0.85 & 17.22 & 5.66 & 5.47 $\pm$ 0.12 & 4.52 $\pm$ 0.11 & 2.73 $\pm$ 0.11 & 2.41 $\pm$ 0.12 & 0.90 $\pm$ 0.11  \\
          & 18.32 & 7.87 & 2.56 $\pm$ 0.24 & 1.98 $\pm$ 0.23 & 1.87 $\pm$ 0.15 & 1.27 $\pm$ 0.15 & --  \\
G0.84+0.18 & 5.89 & 2.72 & 0.37 $\pm$ 0.06 & 0.24 $\pm$ 0.06 & -- & -- & --  \\
G1.15$-$0.12 & $-$16.80 & 2.92 & 0.54 $\pm$ 0.12 & 0.46 $\pm$ 0.13 & -- & -- & --  \\
G5.89$-$0.39 & 9.20 & 4.54 & 2.73 $\pm$ 0.09 & 2.49 $\pm$ 0.09 & 1.48 $\pm$ 0.09 & 1.10 $\pm$ 0.09 & --  \\
G5.90$-$0.43 & 7.39 & 4.87 & 3.39 $\pm$ 0.30 & 2.80 $\pm$ 0.29 & 1.68 $\pm$ 0.30 & 1.01 $\pm$ 0.28 & --  \\
          & 7.23 & 5.12 & 2.75 $\pm$ 0.21 & 1.84 $\pm$ 0.21 & 1.25 $\pm$ 0.20 & 0.96 $\pm$ 0.19 & --  \\
G5.90$-$0.44 & 9.73 & 2.63 & 1.02 $\pm$ 0.08 & 0.66 $\pm$ 0.07 & 0.31 $\pm$ 0.07 & 0.22 $\pm$ 0.07 & --  \\
G8.14+0.23 & 18.60 & 4.34 & 0.91 $\pm$ 0.10 & 0.66 $\pm$ 0.10 & 0.32 $\pm$ 0.10 & 0.29 $\pm$ 0.10 & --  \\
G8.67$-$0.36 & 35.02 & 5.06 & 2.74 $\pm$ 0.10 & 2.48 $\pm$ 0.10 & 1.31 $\pm$ 0.10 & 0.76 $\pm$ 0.09 & 0.24 $\pm$ 0.09  \\
G8.68$-$0.37 & 38.01 & 4.40 & 2.29 $\pm$ 0.09 & 2.25 $\pm$ 0.09 & 0.96 $\pm$ 0.09 & 0.72 $\pm$ 0.09 & 0.26 $\pm$ 0.09  \\
G9.62+0.19 & 4.56 & 6.72 & 3.41 $\pm$ 0.27 & 2.13 $\pm$ 0.25 & 1.59 $\pm$ 0.19 & 1.39 $\pm$ 0.20 & --  \\
G9.99$-$0.03 & 49.56 & 4.48 & 0.34 $\pm$ 0.08 & 0.29 $\pm$ 0.08 & 0.13 $\pm$ 0.07 & -- & --  \\
G10.29$-$0.13 & 13.40 & 4.81 & 2.01 $\pm$ 0.17 & 1.46 $\pm$ 0.17 & 0.68 $\pm$ 0.16 & -- & --  \\
G10.30$-$0.15 & 12.88 & 5.39 & 2.80 $\pm$ 0.37 & 2.39 $\pm$ 0.36 & 0.85 $\pm$ 0.33 & 0.54 $\pm$ 0.32 & --  \\
          & 12.37 & 4.03 & 0.89 $\pm$ 0.14 & 0.77 $\pm$ 0.14 & -- & -- & --  \\
G10.32$-$0.16 & 12.17 & 5.10 & 1.36 $\pm$ 0.21 & 1.13 $\pm$ 0.20 & 0.66 $\pm$ 0.22 & 0.56 $\pm$ 0.22 & --  \\
          & 11.69 & 3.56 & 0.51 $\pm$ 0.06 & 0.39 $\pm$ 0.05 & 0.27 $\pm$ 0.05 & -- & --  \\
G10.34$-$0.14 & 12.07 & 4.55 & 1.24 $\pm$ 0.23 & 0.81 $\pm$ 0.22 & 0.37 $\pm$ 0.20 & 0.23 $\pm$ 0.20 & --  \\
          & 12.51 & 3.92 & 1.44 $\pm$ 0.13 & 1.31 $\pm$ 0.12 & 0.41 $\pm$ 0.12 & -- & --  \\
G10.44$-$0.02 & 73.28 & 4.20 & 0.38 $\pm$ 0.09 & 0.15 $\pm$ 0.09 & -- & -- & --  \\
          & 75.56 & 5.49 & 0.72 $\pm$ 0.15 & 0.75 $\pm$ 0.20 & -- & -- & --  \\
G10.47+0.03 & 67.31 & 11.32 & 7.70 $\pm$ 0.48 & 5.02 $\pm$ 0.45 & 6.89 $\pm$ 0.28 & 6.53 $\pm$ 0.26 & 3.93 $\pm$ 0.25  \\
G10.48+0.03 & 66.34 & 9.03 & 2.40 $\pm$ 0.33 & 1.59 $\pm$ 0.31 & 1.38 $\pm$ 0.18 & 1.00 $\pm$ 0.18 & --  \\
G10.63$-$0.38 & $-$3.69 & 6.69 & 4.26 $\pm$ 0.25 & 3.59 $\pm$ 0.25 & 2.01 $\pm$ 0.18 & 1.62 $\pm$ 0.18 & --  \\
          & $-$4.81 & 6.85 & 2.76 $\pm$ 0.27 & 3.17 $\pm$ 0.28 & 1.29 $\pm$ 0.17 & 1.50 $\pm$ 0.18 & --  \\
G11.94$-$0.62 & 38.11 & 4.24 & 1.74 $\pm$ 0.17 & 1.46 $\pm$ 0.16 & 0.58 $\pm$ 0.15 & -- & --  \\
          & 38.11 & 3.83 & 0.64 $\pm$ 0.13 & 0.41 $\pm$ 0.12 & 0.27 $\pm$ 0.13 & -- & --  \\
G12.68$-$0.18 & 55.99 & 4.13 & 0.86 $\pm$ 0.10 & 0.77 $\pm$ 0.10 & 0.37 $\pm$ 0.09 & -- & --  \\
          & 56.73 & 4.89 & 0.92 $\pm$ 0.18 & 0.81 $\pm$ 0.19 & 0.85 $\pm$ 0.21 & 0.59 $\pm$ 0.19 & --  \\
G12.72$-$0.22 & 34.40 & 5.80 & 0.66 $\pm$ 0.13 & 0.44 $\pm$ 0.13 & 0.21 $\pm$ 0.12 & -- & --  \\
G12.89+0.49 & 33.28 & 4.31 & 1.24 $\pm$ 0.16 & 1.32 $\pm$ 0.16 & 0.78 $\pm$ 0.17 & 1.05 $\pm$ 0.17 & --  \\
G12.91$-$0.26 & 37.40 & 4.18 & 1.07 $\pm$ 0.23 & 0.79 $\pm$ 0.21 & 0.33 $\pm$ 0.20 & -- & --  \\
          & 37.47 & 5.03 & 2.22 $\pm$ 0.16 & 2.65 $\pm$ 0.16 & 1.56 $\pm$ 0.16 & 1.87 $\pm$ 0.17 & --  \\
G14.60+0.02 & 26.40 & 5.43 & 0.48 $\pm$ 0.05 & 0.67 $\pm$ 0.06 & 0.29 $\pm$ 0.05 & 0.26 $\pm$ 0.05 & --  \\
G14.99$-$0.70 & 19.09 & 6.51 & 1.61 $\pm$ 0.29 & 0.90 $\pm$ 0.27 & 0.94 $\pm$ 0.24 & -- & --  \\
G15.03$-$0.68 & 19.99 & 4.38 & 0.67 $\pm$ 0.10 & 0.51 $\pm$ 0.10 & 0.11 $\pm$ 0.10 & -- & --  \\
          & 19.65 & 3.42 & 0.86 $\pm$ 0.09 & 0.84 $\pm$ 0.09 & 0.60 $\pm$ 0.09 & -- & --  \\
G16.59$-$0.05 & 59.91 & 5.00 & 1.21 $\pm$ 0.32 & 1.50 $\pm$ 0.33 & 0.53 $\pm$ 0.31 & -- & --  \\
G16.86$-$2.16 & 18.14 & 5.79 & 1.41 $\pm$ 0.34 & 1.28 $\pm$ 0.33 & 0.49 $\pm$ 0.28 & 0.34 $\pm$ 0.30 & --  \\
          & 17.98 & 4.59 & 1.18 $\pm$ 0.17 & 1.66 $\pm$ 0.18 & 0.67 $\pm$ 0.17 & 0.40 $\pm$ 0.17 & --  \\
G19.36$-$0.03 & 25.83 & 3.87 & 0.55 $\pm$ 0.11 & 0.53 $\pm$ 0.10 & -- & -- & --  \\
G19.47+0.17 & 20.31 & 6.04 & 1.93 $\pm$ 0.25 & 2.01 $\pm$ 0.26 & 0.99 $\pm$ 0.24 & 0.81 $\pm$ 0.23 & --  \\
G19.61$-$0.13 & 57.07 & 5.68 & 0.43 $\pm$ 0.16 & 0.69 $\pm$ 0.17 & -- & -- & --  \\
G22.36+0.07 & 84.31 & 1.62 & 0.17 $\pm$ 0.04 & 0.11 $\pm$ 0.04 & -- & -- & --  \\
G23.44$-$0.18 & 101.64 & 5.43 & 2.18 $\pm$ 0.17 & 1.88 $\pm$ 0.16 & 0.80 $\pm$ 0.15 & 0.83 $\pm$ 0.15 & --  \\
          & 102.48 & 5.45 & 2.45 $\pm$ 0.18 & 2.00 $\pm$ 0.18 & 0.89 $\pm$ 0.16 & 1.06 $\pm$ 0.17 & --  \\
G24.79+0.08 & 110.81 & 6.63 & 3.60 $\pm$ 0.22 & 3.42 $\pm$ 0.21 & 2.38 $\pm$ 0.20 & 1.92 $\pm$ 0.20 & 0.45 $\pm$ 0.18  \\
          & 112.03 & 5.94 & 3.68 $\pm$ 0.22 & 3.58 $\pm$ 0.22 & 2.45 $\pm$ 0.21 & 2.55 $\pm$ 0.21 & --  \\
G25.65+1.05 & 42.55 & 4.55 & 1.11 $\pm$ 0.21 & 0.93 $\pm$ 0.21 & 0.78 $\pm$ 0.23 & -- & --  \\
G25.71+0.04 & 99.01 & 5.26 & 0.45 $\pm$ 0.11 & 0.16 $\pm$ 0.11 & 0.19 $\pm$ 0.09 & -- & --  \\
G25.83$-$0.18 & 93.81 & 5.25 & 2.13 $\pm$ 0.15 & 1.89 $\pm$ 0.15 & 0.92 $\pm$ 0.15 & 0.87 $\pm$ 0.14 & 0.33 $\pm$ 0.14  \\
          & 94.38 & 5.27 & 1.84 $\pm$ 0.17 & 1.88 $\pm$ 0.17 & 0.96 $\pm$ 0.17 & -- & --  \\
G28.15+0.00 & 98.95 & 3.22 & 0.54 $\pm$ 0.12 & 0.49 $\pm$ 0.12 & 0.20 $\pm$ 0.11 & -- & --  \\
G28.20$-$0.05 & 95.86 & 5.83 & 1.19 $\pm$ 0.25 & 1.46 $\pm$ 0.26 & 1.31 $\pm$ 0.25 & 0.59 $\pm$ 0.22 & --  \\
G28.28$-$0.36 & 48.69 & 4.09 & 0.58 $\pm$ 0.11 & 0.41 $\pm$ 0.11 & -- & -- & --  \\
G28.83$-$0.25 & 86.83 & 2.70 & 0.67 $\pm$ 0.15 & 0.46 $\pm$ 0.14 & 0.32 $\pm$ 0.14 & -- & --  \\
	\hline
      \end{tabular}
      \end{minipage}
    \end{center}
\end{table*}
\clearpage
\newpage
\begin{table*}
  \begin{center}
    \begin{minipage}{155mm}
      \contcaption{}
      \begin{tabular}{lccccccc}
	\hline
	Source\,$^{\alpha}$  & V$_{\rm LSR}$\,$^{\beta}$ & $\delta$V & T$_{\rm MB}$~dv,~K\,=\,0 & T$_{\rm MB}$~dv,~K\,=\,1 & T$_{\rm MB}$~dv,~K\,=\,2 & T$_{\rm MB}$~dv,~K\,=\,3 & T$_{\rm MB}$~dv,~K\,=\,4 \\
	& (\kms) & (\kms)      & (K \kms)     & (K \kms)     & (K \kms)     & (K \kms)     & (K \kms)     \\
	\hline
G29.96$-$0.02 & 97.75 & 5.02 & 2.02 $\pm$ 0.26 & 1.52 $\pm$ 0.24 & 1.24 $\pm$ 0.26 & 1.02 $\pm$ 0.26 & 0.24 $\pm$ 0.23  \\
          & 97.12 & 6.53 & 2.21 $\pm$ 0.25 & 1.13 $\pm$ 0.22 & 1.50 $\pm$ 0.20 & 1.50 $\pm$ 0.23 & --  \\
G29.98$-$0.04 & 102.27 & 8.76 & 0.79 $\pm$ 0.10 & 0.46 $\pm$ 0.17 & -- & -- & --  \\
          & 100.49 & 3.02 & 0.71 $\pm$ 0.20 & 0.64 $\pm$ 0.20 & -- & -- & --  \\
G30.71$-$0.06 & 91.19 & 4.60 & 3.36 $\pm$ 0.29 & 3.03 $\pm$ 0.29 & 1.41 $\pm$ 0.28 & 1.35 $\pm$ 0.28 & 0.60 $\pm$ 0.27  \\
          & 91.37 & 4.89 & 2.22 $\pm$ 0.22 & 1.98 $\pm$ 0.21 & 1.28 $\pm$ 0.23 & 1.43 $\pm$ 0.23 & --  \\
G30.76$-$0.05 & 92.65 & 4.18 & 0.67 $\pm$ 0.12 & 0.69 $\pm$ 0.12 & 0.15 $\pm$ 0.12 & -- & --  \\
          & 98.26 & 2.15 & 1.35 $\pm$ 0.33 & 0.68 $\pm$ 0.22 & -- & -- & --  \\
G30.79+0.20 & 81.39 & 4.28 & 0.61 $\pm$ 0.26 & 0.76 $\pm$ 0.26 & 0.29 $\pm$ 0.25 & -- & --  \\
G30.82$-$0.05 & 98.59 & 6.48 & 5.98 $\pm$ 0.44 & 4.82 $\pm$ 0.38 & 2.92 $\pm$ 0.33 & 2.18 $\pm$ 0.32 & 0.73 $\pm$ 0.31  \\
          & 98.10 & 7.24 & 5.48 $\pm$ 0.33 & 3.88 $\pm$ 0.29 & 2.81 $\pm$ 0.23 & 3.36 $\pm$ 0.23 & --  \\
G30.90+0.16 & 105.62 & 4.19 & 0.43 $\pm$ 0.13 & 0.25 $\pm$ 0.12 & -- & -- & --  \\
G31.28+0.06 & 109.47 & 4.11 & 1.12 $\pm$ 0.29 & 1.13 $\pm$ 0.29 & 0.45 $\pm$ 0.28 & -- & --  \\
          & 108.68 & 1.61 & 0.60 $\pm$ 0.12 & 0.32 $\pm$ 0.11 & -- & -- & --  \\
G31.41+0.31 & 97.64 & 7.44 & 2.64 $\pm$ 0.31 & 2.26 $\pm$ 0.29 & 2.56 $\pm$ 0.26 & 2.33 $\pm$ 0.29 & 1.04 $\pm$ 0.25  \\
          & 98.11 & 7.33 & 3.83 $\pm$ 0.24 & 3.35 $\pm$ 0.23 & 3.02 $\pm$ 0.20 & 2.88 $\pm$ 0.21 & --  \\
G316.81$-$0.06 & $-$39.69 & 3.70 & 0.99 $\pm$ 0.14 & 0.83 $\pm$ 0.14 & 0.67 $\pm$ 0.14 & 0.49 $\pm$ 0.14 & --  \\
          & $-$43.70 & 7.44 & 0.24 $\pm$ 0.33 & 1.22 $\pm$ 0.36 & -- & -- & --  \\
G318.95$-$0.20 & $-$34.68 & 4.47 & 1.48 $\pm$ 0.12 & 1.25 $\pm$ 0.12 & 0.76 $\pm$ 0.12 & 0.58 $\pm$ 0.12 & --  \\
G323.74$-$0.26 & $-$50.11 & 4.30 & 1.55 $\pm$ 0.12 & 1.22 $\pm$ 0.11 & 0.82 $\pm$ 0.12 & 0.75 $\pm$ 0.12 & --  \\
          & $-$48.73 & 2.90 & 0.85 $\pm$ 0.15 & 0.64 $\pm$ 0.14 & -- & -- & --  \\
G331.28$-$0.19 & $-$87.85 & 5.36 & 1.60 $\pm$ 0.17 & 1.24 $\pm$ 0.16 & 0.94 $\pm$ 0.14 & 0.81 $\pm$ 0.14 & --  \\
G332.73$-$0.62 & $-$50.74 & 2.55 & 0.37 $\pm$ 0.10 & 0.30 $\pm$ 0.09 & -- & -- & --  \\
\hline
      \end{tabular}
      \begin{flushleft}
	$^{\alpha}$~For each source the parameters of the
	\chthreecn\,(5\,--\,4) fits are presented in the upper row and
	those of the \chthreecn\,(6\,--\,5) fits are presented in the
	lower row.\\
	\smallskip
	$^{\beta}$~Quoted V$_{\rm LSR}$ is referenced to the K\,=\,0 component.
      \end{flushleft}
    \end{minipage}
  \end{center}
\end{table*}

\newpage
\clearpage
\begin{table}
  \begin{center}
    \begin{minipage}{37mm}
      \caption{Sources without detected \chthreecn.}\label{tab:ch3cn_nond}
      \begin{tabular}{lc}
	\hline
	Source & rms-noise\,$^{\alpha}$\\
	       & (mK)\\
	\hline
	G0.21+0.00  & 0.101 \\
	G0.50+0.19  & 0.060 \\
	G2.54+0.20  & 0.149 \\
	G6.54$-$0.11  & 0.075 \\
	G6.61$-$0.08  & 0.108 \\
	G10.10+0.72 & 0.054 \\
	G10.63$-$0.33 & 0.087 \\
	G11.50$-$1.49 & 0.143 \\
	G11.94$-$0.15 & 0.082 \\
	G11.99$-$0.27 & 0.080 \\
	G12.03$-$0.03 & 0.083 \\
	G12.18$-$0.12 & 0.096 \\
	G12.21$-$0.09 & 0.099 \\
	G15.03$-$0.71 & 0.064 \\
	G19.49+0.15 & 0.082 \\
	G19.70$-$0.27 & 0.074 \\
	G21.88+0.01 & 0.036 \\
	G23.26$-$0.24 & 0.093 \\
	G23.71$-$0.20 & 0.088 \\
	G24.85+0.09 & 0.098 \\
	G28.31$-$0.39 & 0.079 \\
	G29.87$-$0.04 & 0.106 \\
	G30.59$-$0.04 & 0.093 \\
	G30.78+0.23 & 0.089 \\
	G30.82+0.28 & 0.064 \\
	\hline
      \end{tabular}
      \begin{flushleft}
	$^{\alpha}$~1\,$\sigma$ noise in the \chthreecn\,(5\,--\,4) spectrum.
      \end{flushleft}
    \end{minipage}
\end{center}
\end{table}

\newpage
\clearpage
\begin{figure*}
  \addtocounter{table}{1}
  \includegraphics[angle=180]{tables/table6a.epsi}
\end{figure*}
\newpage
\clearpage
\begin{figure*}
  \includegraphics[angle=180]{tables/table6b.epsi}
\end{figure*}
\newpage
\clearpage
\begin{table*}
  \begin{center}
  \begin{minipage}{95mm}
    \caption{Gaussian fits to \hcop~line wings and blended lines.}\label{tab:hcop_linewings}
    \begin{tabular}{lcccccccl}
      \hline
      Source\,$^{\alpha}$  & T$_{\rm MB}$~dv & V$_{\rm LSR}$ & $\delta$V & Code\,$^{\beta}$\\
              & (K \kms)    & (\kms)    & (\kms)      \\
      \hline
G0.32$-$0.20 & 9.31 $\pm$ 0.61 & 18.48 $\pm$ 0.24 & 11.88 $\pm$ 0.93 & w \\
G0.50+0.19 & 16.77 $\pm$ 2.11 & 26.02 $\pm$ 2.57 & 43.07 $\pm$ 6.50 & b \\
G0.55$-$0.85 & 9.84 $\pm$ 0.54 & 22.23 $\pm$ 0.09 & 4.77 $\pm$ 0.31 & b \\
G2.54+0.20 & 7.76 $\pm$ 1.04 & 9.76 $\pm$ 0.14 & 5.57 $\pm$ 0.53 & w \\
G5.89$-$0.39 & 47.45 $\pm$ 3.11 & 10.91 $\pm$ 0.49 & 26.75 $\pm$ 1.21 & w \\
G6.61$-$0.08 & 31.45 $\pm$ 0.74 & $-$1.75 $\pm$ 0.27 & 23.98 $\pm$ 0.66 & b \\
G8.67$-$0.36 & 14.61 $\pm$ 1.71 & 35.23 $\pm$ 0.25 & 13.20 $\pm$ 0.91 & w \\
G9.62+0.19 & 18.37 $\pm$ 0.96 & 4.29 $\pm$ 0.14 & 11.47 $\pm$ 0.46 & w \\
G10.30$-$0.15 & 35.38 $\pm$ 0.95 & 11.11 $\pm$ 0.10 & 11.90 $\pm$ 0.24 & w \\
G10.32$-$0.16 & 13.88 $\pm$ 0.76 & 12.23 $\pm$ 0.10 & 7.48 $\pm$ 0.36 & w \\
G10.34$-$0.14 & 9.74 $\pm$ 0.65 & 11.50 $\pm$ 0.22 & 9.95 $\pm$ 0.57 & w \\
G10.44$-$0.02 & 3.17 $\pm$ 0.23 & 65.92 $\pm$ 0.16 & 4.87 $\pm$ 0.46 & b \\
G10.48+0.03 & 15.60 $\pm$ 1.62 & 66.12 $\pm$ 0.21 & 13.80 $\pm$ 0.96 & w \\
G10.63$-$0.38 & 11.26 $\pm$ 1.07 & 3.10 $\pm$ 0.23 & 6.40 $\pm$ 0.42 & b \\
G11.50$-$1.49 & 10.89 $\pm$ 0.74 & 13.64 $\pm$ 0.68 & 20.46 $\pm$ 1.35 & w \\
G11.94$-$0.15 & 1.38 $\pm$ 0.27 & 47.20 $\pm$ 0.21 & 2.66 $\pm$ 0.64 & b \\
G11.99$-$0.27 & 4.76 $\pm$ 0.45 & 59.83 $\pm$ 0.72 & 17.40 $\pm$ 2.38 & w \\
G12.91$-$0.26 & 23.45 $\pm$ 1.28 & 42.54 $\pm$ 0.67 & 27.54 $\pm$ 1.72 & w \\
G14.60+0.02 & 2.04 $\pm$ 0.13 & 37.93 $\pm$ 0.14 & 4.05 $\pm$ 0.32 & b \\
G15.03$-$0.71 & 3.13 $\pm$ 0.98 & 18.50 $\pm$ 1.09 & 8.96 $\pm$ 1.03 & wb \\
G16.86$-$2.16 & 16.41 $\pm$ 0.95 & 18.44 $\pm$ 0.25 & 17.59 $\pm$ 0.83 & w \\
G19.47+0.17 & 2.64 $\pm$ 0.41 & 27.23 $\pm$ 0.24 & 4.07 $\pm$ 0.87 & b \\
G19.49+0.15 & 1.22 $\pm$ 0.26 & 17.71 $\pm$ 0.38 & 3.35 $\pm$ 0.83 & b \\
G19.70$-$0.27 & 1.05 $\pm$ 0.26 & 38.51 $\pm$ 0.37 & 3.32 $\pm$ 1.08 & b \\
G24.79+0.08 & 13.17 $\pm$ 0.98 & 111.37 $\pm$ 0.21 & 17.71 $\pm$ 1.39 & w \\
G24.79+0.08 & 0.43 $\pm$ 0.17 & 102.30 $\pm$ 0.00 & 1.35 $\pm$ 0.62 & b \\
G24.85+0.09 & 3.54 $\pm$ 0.34 & 106.81 $\pm$ 0.55 & 12.36 $\pm$ 1.32 & w \\
G25.65+1.05 & 26.85 $\pm$ 0.72 & 41.34 $\pm$ 0.23 & 21.83 $\pm$ 0.65 & w \\
G25.83$-$0.18 & 10.03 $\pm$ 0.60 & 95.08 $\pm$ 0.63 & 21.70 $\pm$ 2.14 & w \\
G28.83$-$0.25 & 11.14 $\pm$ 0.62 & 89.25 $\pm$ 0.31 & 13.98 $\pm$ 0.77 & w \\
G29.87$-$0.04 & 12.86 $\pm$ 1.06 & 99.72 $\pm$ 0.26 & 11.71 $\pm$ 0.76 & w \\
G29.96$-$0.02 & 6.74 $\pm$ 2.49 & 95.76 $\pm$ 1.91 & 12.24 $\pm$ 2.88 & w \\
G29.96$-$0.02 & 1.86 $\pm$ 1.27 & 103.48 $\pm$ 0.44 & 3.59 $\pm$ 1.37 & b \\
G29.98$-$0.04 & 12.09 $\pm$ 4.09 & 101.74 $\pm$ 1.65 & 17.45 $\pm$ 2.12 & w \\
G29.98$-$0.04 & 14.02 $\pm$ 2.71 & 95.81 $\pm$ 0.15 & 6.72 $\pm$ 0.60 & b \\
G30.71$-$0.06 & 7.74 $\pm$ 1.51 & 99.09 $\pm$ 2.40 & 27.20 $\pm$ 3.69 & w \\
G30.79+0.20 & 3.87 $\pm$ 0.34 & 93.62 $\pm$ 0.33 & 8.34 $\pm$ 0.91 & b \\
G30.82$-$0.05 & 8.22 $\pm$ 1.35 & 104.12 $\pm$ 0.00 & 5.01 $\pm$ 0.71 & b \\
G318.95$-$0.20 & 6.26 $\pm$ 1.00 & $-$31.33 $\pm$ 0.53 & 6.73 $\pm$ 0.68 & w \\
G323.74$-$0.26 & 7.64 $\pm$ 0.69 & $-$48.52 $\pm$ 0.46 & 15.47 $\pm$ 1.97 & w \\
G332.73$-$0.62 & 4.14 $\pm$ 0.33 & $-$56.21 $\pm$ 0.20 & 5.56 $\pm$ 0.54 & b \\
G331.28$-$0.19 & 11.29 $\pm$ 0.72 & $-$89.34 $\pm$ 0.42 & 18.51 $\pm$ 1.26 & w \\
      \hline
    \end{tabular}
  \begin{flushleft}
    $^{\alpha}$~The fits quoted here were {\it subtracted} from the
    \hcop~spectra before analysing the line profile. We assume that
    these blended lines and line wings do not come from the same
    region as the \hthirteencop~emission.\\
    \smallskip
    $^{\beta}$~`w' denotes a Gaussian fit to a \hcop\,(1\,--\,0) line wing, `b' denotes
    a Gaussian fit to a blended \hcop\,(1\,--\,0) line.
  \end{flushleft}
  \end{minipage}
\end{center}
\end{table*}

\newpage
\clearpage
\begin{table*}
  \begin{center}
    \begin{minipage}{173mm}
      \caption{Parameters Derived from \chthreecn.}\label{tab:trot}      
      \begin{tabular}{lcccccclccccc}
	\hline
      Source & T$_{\rm rot}$\,$^{\alpha}$ & Column Density\,$^{\beta}$ & \multicolumn{3}{c}{Notes\,$^{\gamma}$} && Source & T$_{\rm rot}$\,$^{\alpha}$ & Column Density\,$^{\beta}$ & \multicolumn{3}{c}{Notes\,$^{\gamma}$} \\
	     & (K)                    & ($\times 10^{13}$ m$^{-2}$) &           &&                     &&        & (K)                & ($\times 10^{13}$ m$^{-2}$) \\
      \cline{1-6} \cline{8-13}
G0.55$-$0.85 & 60 $\pm$ 3 & 7.2 $\pm$ 2.2 & 5 & M & R && G12.91$-$0.26 & 68 $\pm$ 16 & 2.0 $\pm$ 0.6 & b & M & -- \\
            & 63 $\pm$ 8 & 3.2 $\pm$ 1.0 & 6 &  &  && G14.60+0.02 & 63 $\pm$ 12 & 0.8 $\pm$ 0.2 & 5 & M & R \\
            & 60 $\pm$ 6 & 6.2 $\pm$ 1.9 & b &  &  && G15.03$-$0.68 & 78 $\pm$ 16 & 1.2 $\pm$ 0.4 & 6 & M & -- \\
G5.89$-$0.39 & 56 $\pm$ 4 & 3.5 $\pm$ 1.1 & 5 & -- & R && G16.86$-$2.16 & 36 $\pm$ 16 & 1.1 $\pm$ 0.3 & 5 & M & -- \\
G5.90$-$0.43 & 46 $\pm$ 8 & 3.3 $\pm$ 1.0 & 5 & M & -- &&             & 43 $\pm$ 9 & 1.1 $\pm$ 0.3 & 6 & & \\
            & 45 $\pm$ 6 & 2.1 $\pm$ 0.6 & 6 &  &  &&             & 41 $\pm$ 13 & 1.1 $\pm$ 0.3 & b & & \\
            & 45 $\pm$ 7 & 2.5 $\pm$ 0.8 & b &  &  && G19.47+0.17 & 55 $\pm$ 13 & 2.5 $\pm$ 0.8 & 5 & M & -- \\
G5.90$-$0.44 & 32 $\pm$ 5 & 0.6 $\pm$ 0.2 & 5 & -- & -- && G23.44$-$0.18 & 47 $\pm$ 7 & 2.2 $\pm$ 0.7 & 5 & M & -- \\
G8.14+0.23 & 42 $\pm$ 10 & 0.7 $\pm$ 0.2 & 5 & M & R &&             & 47 $\pm$ 6 & 2.0 $\pm$ 0.6 & 6 & & \\
G8.67$-$0.36 & 43 $\pm$ 3 & 2.6 $\pm$ 0.8 & 5 & M & R &&             & 47 $\pm$ 7 & 2.1 $\pm$ 0.6 & b & & \\
G8.68$-$0.37 & 44 $\pm$ 3 & 2.2 $\pm$ 0.7 & 5 & M & -- && G24.79+0.08 & 74 $\pm$ 9 & 6.7 $\pm$ 2.0 & 5 & M & -- \\
G9.62+0.19 & 60 $\pm$ 9 & 4.1 $\pm$ 1.2 & 5 & M & R &&             & 81 $\pm$ 9 & 6.3 $\pm$ 1.9 & 6 & & \\
G9.99$-$0.03 & 38 $\pm$ 29 & 0.3 $\pm$ 0.1 & 5 & M & -- &&             & 77 $\pm$ 9 & 6.5 $\pm$ 1.9 & b & & \\
G10.29$-$0.13 & 30 $\pm$ 8 & 1.2 $\pm$ 0.4 & 5 & M & -- && G25.83$-$0.18 & 53 $\pm$ 8 & 2.5 $\pm$ 0.8 & 5 & M & -- \\
G10.30$-$0.15 & 32 $\pm$ 8 & 1.8 $\pm$ 0.5 & 5 & M & R &&             & 55 $\pm$ 19 & 2.0 $\pm$ 0.6 & 6 & & \\
G10.32$-$0.16 & 56 $\pm$ 19 & 1.6 $\pm$ 0.5 & 5 & M & -- &&             & 53 $\pm$ 15 & 2.3 $\pm$ 0.7 & b & & \\
            & 53 $\pm$ 24 & 0.5 $\pm$ 0.1 & 6 &  &  && G28.20$-$0.05 & 74 $\pm$ 26 & 2.6 $\pm$ 0.8 & 5 & M & R \\
            & 55 $\pm$ 22 & 0.6 $\pm$ 0.2 & b &  &  && G29.96$-$0.02 & 75 $\pm$ 23 & 3.5 $\pm$ 1.1 & 5 & M & R \\
G10.34$-$0.14 & 30 $\pm$ 12 & 0.7 $\pm$ 0.2 & 5 & M & -- &&             & 116 $\pm$ 39 & 4.8 $\pm$ 1.4 & 6 & & \\
            & 28 $\pm$ 7 & 0.8 $\pm$ 0.2 & 6 &  &  &&             & 86 $\pm$ 32 & 3.9 $\pm$ 1.2 & b & & \\
            & 28 $\pm$ 10 & 0.8 $\pm$ 0.2 & b &  &  && G30.71$-$0.06 & 51 $\pm$ 9 & 3.7 $\pm$ 1.1 & 5 & M & R \\
G10.48+0.03 & 64 $\pm$ 13 & 3.3 $\pm$ 1.0 & 5 & M & -- &&             & 75 $\pm$ 16 & 3.3 $\pm$ 1.0 & 6 & & \\
G10.63$-$0.38 & 52 $\pm$ 5 & 4.8 $\pm$ 1.4 & 5 & M & R &&             & 57 $\pm$ 13 & 3.5 $\pm$ 1.1 & b & & \\
            & 55 $\pm$ 7 & 3.1 $\pm$ 0.9 & 6 &  &  && G30.76$-$0.05 & 29 $\pm$ 14 & 0.4 $\pm$ 0.1 & 5 & M & -- \\
            & 53 $\pm$ 6 & 3.9 $\pm$ 1.2 & b &  &  && G30.82$-$0.05 & 52 $\pm$ 6 & 6.6 $\pm$ 2.0 & 5 & M & -- \\
G11.94$-$0.62 & 32 $\pm$ 9 & 1.1 $\pm$ 0.3 & 5 & M & R &&             & 78 $\pm$ 9 & 7.7 $\pm$ 2.3 & 6 & & \\
G12.68$-$0.18 & 44 $\pm$ 18 & 0.8 $\pm$ 0.2 & 5 & M & -- &&             & 61 $\pm$ 8 & 7.1 $\pm$ 2.1 & b & & \\
            & 84 $\pm$ 35 & 1.6 $\pm$ 0.5 & 6 &  &  && G31.41+0.31 & 98 $\pm$ 13 & 8.3 $\pm$ 2.5 & 6 & M & R \\
            & 52 $\pm$ 28 & 0.9 $\pm$ 0.3 & b &  &  && G316.81$-$0.06 & 73 $\pm$ 24 & 1.7 $\pm$ 0.5 & 5 & M & -- \\
G12.72$-$0.22 & 28 $\pm$ 17 & 0.4 $\pm$ 0.1 & 5 & -- & -- && G318.95$-$0.20 & 55 $\pm$ 9 & 1.8 $\pm$ 0.5 & 5 & M & -- \\
G12.89+0.49 & 131 $\pm$ 53 & 5.2 $\pm$ 1.6 & 5 & M & -- && G323.74$-$0.26 & 67 $\pm$ 12 & 2.3 $\pm$ 0.7 & 5 & M & -- \\
G12.91$-$0.26 & 28 $\pm$ 18 & 0.6 $\pm$ 0.2 & 5 & M & -- && G331.28$-$0.19 & 74 $\pm$ 16 & 2.7 $\pm$ 0.8 & 5 & M & -- \\
            & 94 $\pm$ 15 & 5.0 $\pm$ 1.5 & 6 \\
\hline
      \end{tabular}
      \begin{flushleft}
	$^{\alpha}$ Quoted errors in T$_{\rm rot}$ are the 1\,$\sigma$
	uncertainties arising from the least-squares straight line fit to
	the rotational diagram.\\ 
	\smallskip
	$^{\beta}$ Quoted errors in N are $\sim$\,30 \%, arising from the
	relative calibration error between observations.\\
	\smallskip
	$^{\gamma}$ Results marked with a `6' or a `5' were derived from
	individual J\,=\,(6\,--\,5) or J\,=\,(5\,--\,4) rotational ladders
	respectively. Where we could derive T$_{\rm rot}$ and N from both
	J-transitions we calculated the weighted average, marked with a
	`b' in the notes column. `R' denotes associated 8.6\,GHz radio emission
	and `M' marks a methanol maser site.\\
      \end{flushleft} 
    \end{minipage}
  \end{center}
\end{table*}

\newpage
\clearpage
\begin{table*}
  \begin{center}
    \begin{minipage}{175mm}
      \caption{Comparison of \chthreecn~Parameters to Previous Surveys}\label{tab:comparison}
      \begin{tabular}{lcccccc}
	\hline
	& This Paper & \cite{Pankonin2001} & \cite{Hatchell1998} & \cite{Olmi1993} & \cite{Churchwell1992}\\
	Antenna Diameter: &  22-m & 10-m & 15-m & 30-m & 30-m\\
	Beam Size:        &  35\,\arcsec & 35\,\arcsec & 21\,\arcsec,\,14\,\arcsec & 21\,\arcsec,\,17\,\arcsec,\,12\,\arcsec & 25\,\arcsec,\,11\,\arcsec\\
	(Assumed) source size: & 35\,\arcsec & 35\,\arcsec & $<$\,1.5\,\arcsec & 5\,\arcsec & 10\,\arcsec\\
	\hline
	& \multicolumn{5}{c}{Rotational Temperature, T$_{\rm rot}$\,(K)}\\
	Transitions Used: &(5\,--\,4),\,(6\,--\,5) & (12\,--\,11) & (13\,--\,12) and (19\,--\,18) & (6\,--\,5),\,(8\,--\,7),\,(12\,--\,11) & (6\,--\,5),\,(12\,--\,11)\\
	\hline
	G5.89-0.39  & 56  & 75  & d            & --  & 84  \\
	G9.62+0.19  & 60  & 135 & 119 and 65   & 123 & --  \\
	G10.30-0.15 & 32  & n   & n            & 34  & --  \\
	G10.47+0.03 & d   & 196 & 87 and 134   & d   & --  \\
	G10.63-0.38 & 53  & 103 & --           & --  & 102 \\
	G11.94-0.62 & 32  & n   & --           & --  & d   \\
	G12.21-0.09 & n   & d   & d            & 72  & --  \\
	G15.03-0.68 & 78  & n   & --           & 52  & --  \\
	G25.65+1.05 & 166 & 197 & --           & --  & --  \\
	G29.96-0.02 & 86  & 150 & 114 and 141  & --  & 144 \\
	G31.41+0.31 & 98  & 292 & 149 and 142  & --  & 372 \\
	\hline      
	& \multicolumn{5}{c}{Column Density, N$_{\rm col}$\,($\times$10$^{13}$\,cm$^{-2}$)}\\
	Transitions Used: &(5\,--\,4),\,(6\,--\,5) & (12\,--\,11) & (13\,--\,12) & (6\,--\,5),\,(8\,--\,7),\,(12\,--\,11) & (6\,--\,5),\,(12\,--\,11)\\
	\hline
	G5.89-0.39  & 3.5 & 2.3  & d  & --   & 13 \\
	G9.62+0.19  & 4.1 & 1.8  & 2  & 10.6 & -- \\
	G10.30-0.15 & 1.8 & n    & n  & 11.9 & -- \\
	G10.47+0.03 & d   & 8    & 13 & d    & -- \\
	G10.63-0.38 & 3.9 & 2.2  & -- & --   & 11 \\
	G11.94-0.62 & 1.2 & n    & -- & --   & d  \\
	G12.21-0.09 & n   & d    & d  & 7.1  & -- \\
	G15.03-0.68 & 0.8 & n    & -- & 2.6  & -- \\
	G25.65+1.05 & d   & 0.74 & -- & --   & -- \\
	G29.96-0.02 & 3.9 & 1.8  & 3  & --   & 6  \\
	G31.41+0.31 & 8.3 & 8    & 5  & --   & 17 \\
	\hline      
	\hline
      \end{tabular}
      \begin{flushleft}
	d = \chthreecn~detected but no meaningful rotational temperature or column-density could be derived.\\
	\smallskip
	n = \chthreecn~was not detected.\\
	\smallskip
	-- = Source not observed.\\
      \end{flushleft}
    \end{minipage}
  \end{center}
\end{table*}

\newpage
\clearpage
\begin{table*}
  \begin{center}
    \begin{minipage}{175mm}
      \caption{Parameters derived from analysis of the \hcop~and \hthirteencop~lines.}\label{tab:hcop_deriv}
      \begin{tabular}{lcccclccc}
	\hline
	Source & $\tau$\,(\hthirteencop)\,$^{\alpha}$ & N\,(\hthirteencop) & N\,(\hcop)\,$^{\beta}$ && Source & $\tau$\,(\hthirteencop)\,$^{\alpha}$ & N\,(\hthirteencop) & N\,(\hcop)\,$^{\beta}$\\
	       & & ($\times$10$^{13}$\,cm$^{-2}$) & ($\times$10$^{14}$\,cm$^{-2}$) && & & ($\times$10$^{13}$\,cm$^{-2}$) & ($\times$10$^{14}$\,cm$^{-2}$)\\
	\cline{1-4}  \cline{6-9}
	G0.32$-$0.20 & 0.050 & 2.0 & 10.1 && G16.86$-$2.16 & 0.130 & 7.0 & 34.9 \\
G0.50+0.19 & 0.040 & 2.4 & 12.1 && G19.36$-$0.03 & 0.040 & 1.9 & 9.6 \\
G0.55$-$0.85 & 0.190 & 10.2 & 50.8 && G19.47+0.17 & 0.070 & 4.9 & 24.5 \\
G0.84+0.18 & 0.010 & 0.6 & 3.0 && G19.49+0.15 & 0.020 & 0.9 & 4.5 \\
G1.15$-$0.12 & 0.020 & 1.3 & 6.7 && G19.61$-$0.13 & 0.010 & 0.5 & 2.5 \\
G2.54+0.20 & 0.100 & 4.1 & 20.4 && G19.70$-$0.27 & 0.030 & 1.1 & 5.4 \\
G5.89$-$0.39 & 0.150 & 9.4 & 47.1 && G21.88+0.01 & 0.020 & 1.1 & 5.3 \\
G5.90$-$0.43 & 0.100 & 5.1 & 25.7 && G22.36+0.07 & 0.030 & 1.0 & 4.9 \\
G5.90$-$0.44 & 0.070 & 2.0 & 10.2 && G23.26$-$0.24 & 0.030 & 1.2 & 6.2 \\
G6.61$-$0.08 & 0.020 & 0.7 & 3.5 && G23.44$-$0.18 & 0.070 & 4.3 & 21.7 \\
G8.14+0.23 & 0.030 & 1.9 & 9.3 && G23.71$-$0.20 & 0.020 & 1.1 & 5.7 \\
G8.67$-$0.36 & 0.120 & 7.4 & 37.2 && G24.79+0.08 & 0.090 & 5.5 & 27.6 \\
G8.68$-$0.37 & 0.090 & 7.0 & 34.8 && G24.85+0.09 & 0.010 & 0.5 & 2.6 \\
G9.62+0.19 & 0.060 & 4.5 & 22.3 && G25.65+1.05 & 0.090 & 5.1 & 25.3 \\
G9.99$-$0.03 & 0.030 & 1.0 & 4.8 && G25.71+0.04 & 0.030 & 1.3 & 6.5 \\
G10.29$-$0.13 & 0.050 & 3.0 & 14.7 && G25.83$-$0.18 & 0.050 & 2.3 & 11.6 \\
G10.30$-$0.15 & 0.100 & 6.7 & 33.5 && G28.15+0.00 & 0.040 & 1.4 & 6.8 \\
G10.32$-$0.16 & 0.060 & 2.3 & 11.7 && G28.20$-$0.05 & 0.060 & 3.3 & 16.6 \\
G10.34$-$0.14 & 0.060 & 2.8 & 14.2 && G28.28$-$0.36 & 0.020 & 1.1 & 5.4 \\
G10.44$-$0.02 & 0.030 & 2.0 & 10.0 && G28.31$-$0.39 & 0.030 & 1.8 & 9.1 \\
G10.47+0.03 & 0.060 & 6.1 & 30.4 && G28.83$-$0.25 & 0.050 & 2.6 & 13.2 \\
G10.48+0.03 & 0.040 & 4.2 & 20.8 && G29.87$-$0.04 & 0.040 & 1.8 & 9.0 \\
G10.63$-$0.33 & 0.060 & 2.7 & 13.4 && G29.96$-$0.02 & 0.080 & 3.6 & 17.9 \\
G10.63$-$0.38 & 0.160 & 14.3 & 71.4 && G29.98$-$0.04 & 0.060 & 2.2 & 11.1 \\
G11.50$-$1.49 & 0.050 & 1.4 & 7.1 && G30.59$-$0.04 & 0.040 & 2.1 & 10.6 \\
G11.94$-$0.15 & 0.030 & 1.9 & 9.3 && G30.71$-$0.06 & 0.100 & 8.0 & 40.1 \\
G11.94$-$0.62 & 0.070 & 4.4 & 22.2 && G30.76$-$0.05 & 0.040 & 3.6 & 17.8 \\
G11.99$-$0.27 & 0.010 & 0.8 & 3.9 && G30.78+0.23 & 0.050 & 1.1 & 5.5 \\
G12.03$-$0.03 & 0.020 & 0.8 & 4.2 && G30.79+0.20 & 0.030 & 1.3 & 6.7 \\
G12.18$-$0.12 & 0.010 & 1.5 & 7.5 && G30.82$-$0.05 & 0.080 & 5.1 & 25.5 \\
G12.21$-$0.09 & 0.030 & 1.2 & 6.2 && G30.82+0.28 & 0.020 & 0.5 & 2.6 \\
G12.68$-$0.18 & 0.030 & 1.4 & 7.0 && G30.90+0.16 & 0.040 & 1.8 & 8.9 \\
G12.72$-$0.22 & 0.080 & 3.3 & 16.3 && G31.28+0.06 & 0.060 & 4.0 & 19.9 \\
G12.89+0.49 & 0.060 & 2.8 & 13.7 && G31.41+0.31 & 0.030 & 1.8 & 8.9 \\
G12.91$-$0.26 & 0.110 & 6.1 & 30.4 && G316.81$-$0.06 & 0.050 & 3.3 & 16.7 \\
G14.60+0.02 & 0.040 & 3.8 & 19.2 && G318.95$-$0.20 & 0.110 & 4.5 & 22.5 \\
G14.99$-$0.70 & 0.040 & 2.7 & 13.3 && G323.74$-$0.26 & 0.080 & 3.2 & 15.9 \\
G15.03$-$0.68 & 0.160 & 6.8 & 34.1 && G331.28$-$0.19 & 0.050 & 2.9 & 14.7 \\
G15.03$-$0.71 & 0.010 & 0.8 & 3.8 && G332.73$-$0.62 & 0.040 & 0.9 & 4.4 \\
G16.59$-$0.05 & 0.060 & 2.8 & 13.9 \\
	\hline	
      \end{tabular}
      \begin{flushleft}
	$^{\alpha}$~We assumed an excitation temperature of 15\,K when
	calculating the optical depth of the \hthirteencop\,(1\,--\,0)
	line. Excitation temparatures derived from the \hcop\,(1\,--\,0)
	line were anomalously low, most likely as a result of beam
	dilution.\\
	\smallskip
	$^{\beta}$~A Galactic gradient in the $^{12}$C\,/\,$^{13}$C
	isotope ratio has been observed, ranging from 20 to
	70, however, there is considerable uncertainty in these
	results. The [\hcop]\,/\,[\hthirteencop] ratio may also be
	affected by chemistry. We prefer to use an average value of 50
	when calculating the column density of \hcop.
      \end{flushleft}
    \end{minipage}
  \end{center}
\end{table*}

\newpage
\clearpage
\begin{table*}
  \begin{center} 
    \begin{minipage}{155mm}
      \caption{Parameters of the MSX red and MSX dark sources.}\label{tab:msx_red_dark}
      \begin{tabular}{lcccccccccc}
	\hline
	Source       & \chthreecn & \hcop & \hthirteencop & 21\,\micron & 8\,\micron & Maser & Radio & Profile & Wings & T$_{\rm rot}$\\
	\hline
	& \multicolumn{10}{c}{MSX-red}\\
	\cline{2-11}\\	
	G12.18-0.12  & n & y & y & y & n & y & n & --   & -- & -- \\
	G12.21-0.09  & n & y & y & y & n & y & n & blue & -- & -- \\
	G14.60+0.02  & y & y & y & y & n & y & n & --   & -- & 63 \\
	G24.79+0.08  & y & y & y & y & n & y & n & red  & y  & 77 \\
	G25.71+0.04  & y & y & y & y & n & y & n & --   & -- & -- \\
	G28.83-0.25  & y & y & y & y & n & y & n & red  & y  & -- \\
	G30.79+0.20  & y & y & y & y & d & y & n & --   & -- & -- \\
	G323.74-0.26 & y & y & y & y & n & y & n & red  & y  & 67 \\\\
	& \multicolumn{10}{c}{MSX-dark}\\
	\cline{2-11}\\
	G0.26+0.01   & y & y & y & d & d & n & n & --   & -- & -- \\
	G2.54+0.20   & n & y & y & n & d & y & n & blue & y  & -- \\
	G8.68-0.37   & y & y & y & n & d & y & n & blue & -- & 43 \\
	G10.44-0.02  & y & y & y & n & d & y & n & blue & -- & -- \\
	G10.48+0.03  & y & y & y & n & d & y & n & --   & y  & 64 \\
	G14.99-0.70  & y & y & y & n & d & n & n & --   & -- & -- \\
	G23.44-0.18  & y & y & y & n & d & y & n & red  & -- & 47 \\
	G25.83-0.18  & y & y & y & n & d & y & n & blue & y  & 53 \\
	G29.98-0.04  & y & y & y & n & d & y & n & --   & y  & -- \\
	G30.71-0.06  & y & y & y & d & d & y & b & blue & y  & 57 \\
	G30.82-0.05  & y & y & y & n & d & y & n & blue & -- & 61 \\
	G332.73-0.62 & y & y & y & d & d & y & n & --   & -- & -- \\
	\hline
      \end{tabular}
    \end{minipage}
  \end{center}
\end{table*}


\label{lastpage}
\end{document}